\documentclass[%
 reprint,
superscriptaddress,
 amsmath,amssymb,
 aps,
pra,
]{revtex4-2}
\usepackage{graphicx}
\usepackage{dcolumn}
\usepackage{bm}
\usepackage{color}

\usepackage[colorlinks=true,
            urlcolor=blue,linkcolor=blue,
            citecolor=blue,anchorcolor=blue,
            breaklinks=true]{hyperref}
\usepackage{hyperref}

\usepackage[normalem]{ulem}

\begin{document}

\preprint{APS/123-QED}
\title{Inner Energy Relaxation and Growth of Nano-Size Particles}

\author{M. Bredice}
 \email{mitchell.bredice@uconn.edu}
 \affiliation{Department of Physics, University of Connecticut, Storrs, CT 06269, USA}
\author{M. G. Rozman}
 \affiliation{Department of Physics, University of Connecticut, Storrs, CT 06269, USA}
\author{J. Smucker}
 \affiliation{Department of Physics, University of Connecticut, Storrs, CT 06269, USA}
\author{E. Farmer}
 \affiliation{Department of Physics, University of Connecticut, Storrs, CT 06269, USA}
\author{R.C\^{o}t\'{e}}
   \affiliation{UMass-Boston, Department of Physics, Boston, MA 02125, USA}
\author{V. Kharchenko}
 \affiliation{Department of Physics, University of Connecticut, Storrs, CT 06269, USA}
 \affiliation{ITAMP, Center for Astrophysics $|$ Harvard \& Smithsonian, Cambridge, MA 02138, USA}
 \date{\today}
 \begin{abstract}
  In this study, molecular dynamics simulations were conducted to investigate the relaxation of the internal energy in nano-sized particles and its impact on the nucleation of atomic clusters. Quantum-mechanical potentials were utilized to analyze the growth and collision relaxation of the internal energy of Ar$_n$H$^+$ clusters in a metastable Ar gas. The results revealed that small nano-clusters are formed in highly excited rotational-vibrational states, and the relaxation of internal energy and growth of these nascent clusters are concurrent processes with a strong mutual influence. Under non-equilibrium growth conditions, the relaxation of internal energy can delay the cluster growth process. The rates of cluster growth and internal energy relaxation were found to be influenced by energy-transfer collisions between cluster particles and free Ar atoms of the bath gas. Furthermore, the non-equilibrium growth and internal energy relaxation of small nano-clusters were found to depend on the structure of the cluster's atomic shells. An ensemble of molecular dynamics simulations were conducted to investigate the growth, time-evolution of kinetic and total energies of Ar$_n$H$^+$ clusters with specified $n \leq 11$, and the results were explained by collisional relaxation processes described by the Boltzmann equation. Finally, the general relationship between the rates of internal energy relaxation and non-equilibrium growth of nano-particles is discussed.
 \end{abstract}
 \maketitle
 \section{Introduction}
 The nucleation of a new phase from a gas or liquid has been studied for many years across a range of fields, including physics, chemistry, astrophysics, and planetary science. For example, the analysis of the formation of nano-size ice and dust particles is critical for determining their spectral properties in the upper atmospheres of exoplanets \cite{GRUNDY2018232,10.1093/mnras/stz2655,Gao2020,acp-18-4519-2018,Egorov2018,Protonated_water_exp} or in circumstellar disks \cite{2000A&A...362.1127K,Dust_dist_Krivov,doi:10.1098/rsta.2016.0254}. Classical Nucleation Theory \cite{CNTreview,ReviewCNT1,ReviewCNT2,LLkinetics} has been relatively successful in analyzing nucleation processes in mesoscopic and large size systems under thermal equilibrium conditions, but is not suitable for describing the non-equilibrium nucleation of nano-particles and clusters. Theoretical modeling of the onset of the nucleation process is challenging when nano-particles grow under local non-equilibrium and non-homogeneous conditions. Detailed knowledge of the relaxation processes in nano-scale objects is needed to understand how nano-particles evolve toward their equilibrium state. Initial stages of nucleation must be considered as local exothermic processes due to the significant release of kinetic energy during the growth of a new phase. As an illustration of this general phenomenon, we will consider the nucleation of nano-size atomic clusters in Ar$_n$H$^{+}$ + Ar $\rightarrow$ Ar$_{n+1}$\textsuperscript{*}H$^{+}$ transitions. As a result, small nano-particles are formed in highly excited rotational-vibrational states, and their internal energy relaxation occurs alongside cluster growth. The processes of cluster energy relaxation and nucleation strongly influence each other, creating a complex picture of nano-particle growth. Accurate knowledge of inter-particle interactions is necessary to describe possible relaxation mechanisms and their effects on the rate of formation of the new phase, the cluster particles.

Previous investigations of atomic clusters have demonstrated that molecular dynamics (MD) simulations with accurate potentials of inter-particle interactions can describe both non-equilibrium and equilibrium processes of the nucleation of nano-size particles. In the past there has been a significant amount of research utilizing MD methods to study the nucleation of various clusters. For instance, some investigations have focused on pure Ar clusters; specifically on their associated structure\cite{PhysRevB.77.125434}, phase transitions\cite{doi:10.1063/1.469470,REY1992273}, and kinetics of their nucleation\cite{LJMD,ArMD}. In addition, there has been recent investigations of the phase transitions of protonated Ar clusters using classical MD simulations\cite{Oliver}. Moreover, there has been a large focus on the nucleation of clusters composed of molecules commonly found in atmospheres, including the growth of water/ice clusters \cite{WaterHeteroHomoNuc,WaterHeteroNucleation,WaterNucTIP,WaterIon}, and the study of the structure of aerosol nanoparticles \cite{Aerosol}. In the past there also has been some theoretical analyses of small Ar$_n$H$^+$ and Lennard-Jones (LJ) clusters. The focus of those investigations were on the respective ground state structures and explanation of "magic numbers" associated with experimental data \cite{PhysRevA.98.022519, Ar_n_H+_QM,doi:10.1063/1.1485956,Ar2_3_structure, Vafayi2015}. There also is a large amount of laboratory research using supersonic beam experiments involving different species of nanoparticle-sized clusters; where those experiments investigate the growth and abundance of nanoparticle-sized clusters with different numbers of atoms \cite{PhysRevA.98.022519,Schobel2011,Pure_He_Doped_Kr, Kuhn2016,PhysRevLett.51.1538,doi:10.1063/1.458275,PhysRevLett.53.2390,HARRIS1986316,MARK1987245,doi:10.1063/1.456898,doi:10.1063/1.457464,Gatchell2019}. However, these investigations did not consider that small nascent clusters are formed in highly excited states and this establishes the relationship between non-equilibrium cluster growth and its internal energy relaxation.

In this study, MD simulations of the growth of Ar$_n$H$^+$ clusters initiated in the Ar gas by H$^+$ ions were performed. The results showed that nanoparticles of noble gas atoms, seeded by an ion, are initially formed in highly-excited rotational and vibrational states. The energy of the cluster, both the center of mass and internal components, undergoes relaxation through collisions with atoms of the ambient Ar gas. The relaxation of internal energy significantly influences the growth of nano-sizeclusters. This type of effect has previously been observed in the nucleation of metallic nanoparticles \cite{Korenchenko2016}.

The growth of clusters in a bath gas have a large number of relaxation pathways, with energy relaxation being a fundamental one \cite{Koloren__2016, PhysRevE.60.3701}. The mechanisms and rates of energy relaxation for the cluster's translational or internal degrees of freedom can be quite different. In relatively dense ambient gas, the initial translational energy of nascent clusters is efficiently dissipated through multiple collisions with free atoms. The relaxation of the internal kinetic energy of growing clusters can include several pathways, such as the transfer of a fraction of the internal energy in atom-cluster cooling collisions or detachment processes, in which the energetic atoms leave the cluster and significantly reduces the cluster's internal energy. For example, the decay of metastable Ar$^*_n$H$^+$ cluster, which was formed in highly excited rotational-vibrational states, can significantly reduce its internal energy through the auto-detachment process Ar$^*_n$H$^+ \rightarrow$ Ar$_{n-1}$H$^+$ + Ar +$\Delta \epsilon$, where $\Delta \epsilon$ is the kinetic energy of the ejected Ar atom. This process can be viewed as an analogy to the Auger decay of highly excited many-electron atoms \cite{Koloren__2016}. 

The relaxation of hot particles due to energy transfer collisions with a bath gas has been extensively studied for atomic and molecular species \cite{AndersonHardSphere,HydrogenRelaxation,NitrogenRelaxation,SXeRelaxation,KHARCHENKOFastnNitrogen}. The large number of inelastic collision channels, which lead to excitation of the cluster's internal degrees of freedom, complicates the theoretical description of the energy relaxation of small clusters. As a result, some studies have focused on the post-collision relaxation of clusters that are excited by a single collision \cite{AtomicCollisionRelaxation,WaterCollision}. In dilute environments, radiation cooling and heating mechanisms can become important, but are neglected in our investigations because collision relaxation occurs much faster.

The fundamental problem in the physics of nano-particle nucleation is to establish general rules that govern the kinetics of cluster formation and growth under non-equilibrium conditions. The rates of relaxation of inner cluster energy and cluster growth influence each other and both strongly depend on the local parameters of the surrounding gas. Detailed analysis of the rates of atomic sticking and detachment processes in collisions with nano-size clusters could provide insight into predicting the cluster size-distribution under non-equilibrium and equilibrium conditions. In high-density environments (solid/liquid), the energy relaxation may be dominated by many-body/bulk effects. However, in the gas environment, the processes of relaxation of internal cluster energy are relatively simple due to the majority of collisions being single atom-cluster collisions. These collisions simultaneously regulate the rates of cluster cooling and growth.

\section{Simulation Details}
In this study, we used molecular dynamics (MD) simulations to investigate how the internal energy relaxation of nascent nano-clusters affects the nucleation process. We performed the simulations using the Large Atomic/Molecular Massively Parallel Simulator (LAMMPS)\cite{LAMMPS}. The binary Ar-Ar and Ar-H$^+$ interactions are described in detail in \cite{Oliver}. Each simulation began with the random generation of coordinates of 1000 Ar atoms and a single H$^+$ ion, with a minimum distance of 3 \AA\  between each particle. The velocities of the Ar atoms were initialized using the LAMMPS "create velocity" function, while the initial H$^+$ velocity for convenience was set to zero. In our simulations, the initial velocity distributions of the Ar atoms had Gaussian shape with the average kinetic energy that corresponded to the desired temperature. We performed simulations at two temperatures,below and above the Ar boiling point, 40K and 90K respectively. These temperatures and density of the Ar buffer gas were chosen so that the Ar$_n$H$^+$ clusters can grow to sizes $n$ $\sim$ 10 within a few nanoseconds and that allows a lower computational time for each individual simulation. We ran 500 simulations with unique initial conditions for each temperature, with a total simulation time of 2.5 ns. In our previous work, we found that small clusters grew relatively quickly, in $\sim$2-5ns, at a temperature of 90K \cite{OurArticle}. The rate of nucleation of nano-clusters was significantly increased at lower temperatures, such as 40K.

All simulations were performed with a timestep of 1 fs at the Ar atom density of 10$^{20}$ cm$^{-3}$, using the canonical (NVT) ensemble with a Nose-Hoover thermostat and a temperature damping timescale of 100 fs. The MD trajectories were recorded at 1 ps intervals. This arrangement of simulations provides the necessary data for the analysis of the dynamics of the internal degrees of freedom and the motion of the center of mass for each cluster. To analyze the cluster parameters, we used the DBSCAN algorithm implemented in the Clustering.jl package of the Julia programming language. A more detailed discussion of the cluster definition and DBSCAN parameters can be found in \cite{OurArticle}. However, for each cluster its total energy is calculated to ensure it is negative to indicate a true bound cluster.
\section{\texorpdfstring{Formation of Nascent $Ar_nH^+$ Clusters}{Section: Nascent Clusters}}
The current MD simulations have three specific goals. The first is to show that small Ar$_n$H$^+$ clusters are formed in highly excited rotational-vibrational states or, the same, in highly-excited geometrical configurations. The second goal is to determine the relationship between the rate of cluster growth and the rate of internal energy relaxation. The third is to establish how the cooling and growth rates of small nano-clusters depend on the parameters of the ambient gas. The simulations carried out in this work can clarify the dynamics of non-equilibrium cooling of internal degrees of freedom in clusters of different sizes. At low densities of the ambient gas, the cooling of small nano-clusters may be a "bottleneck" in the entire nucleation process and must be "stepped through" before a cluster can grow.

To simulate the growth of independent clusters, only one H$^+$ ion is present in the simulation box, allowing for the formation and tracking of a single Ar$_{n(t)}$H$^+$ cluster in the simulation time from each MD trajectory. This allows one to investigate the evolution of a single independent cluster and to avoid the additional complications arising due to  simultaneous nucleation of many clusters around individual protons, where competition between the growth of different clusters leads to the coalescence regime \cite{LLkinetics, OurArticle}. 

\begin{figure}[ht]
    \centering
    \includegraphics[scale=0.266]{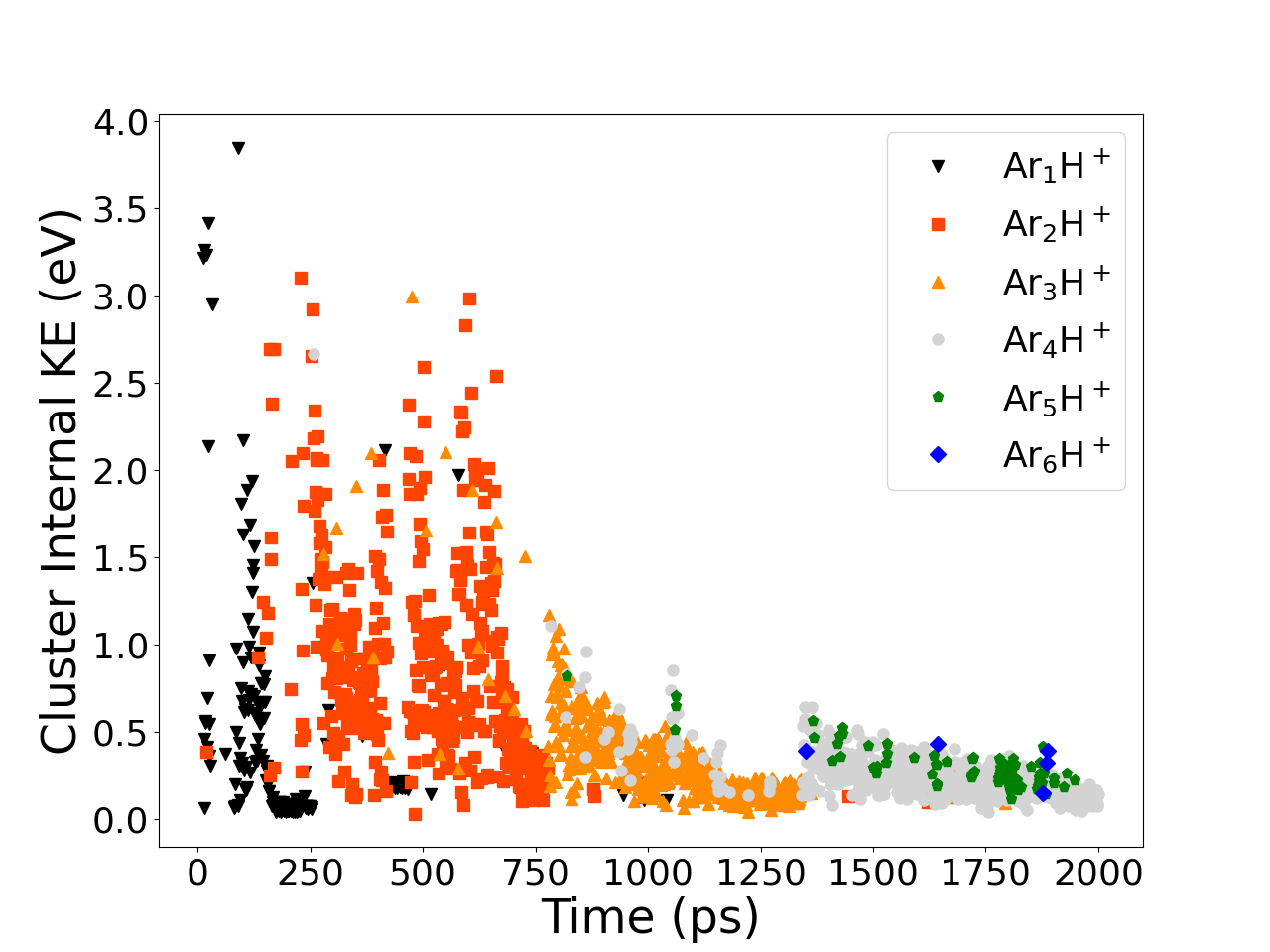}
    \caption{Shown is the internal kinetic energy for an Ar$_{n(t)}$H$^+$ cluster ($n\leq 6$), from a single MD simulation trajectory, where the simulation was performed at 90K. Different marker types and colors indicate changes in cluster size. It is apparent in the first $n\rightarrow n+1$ size transitions that the nascent clusters have a large amount of internal energy, which decreases before subsequent {\it n-}growth. The $n=4$ clusters (gray points) exhibit stability by remaining present at long time scale.} 
    \label{fig:SingleClusterKE}
\end{figure}

The cluster's internal kinetic energy represents the sum of the kinetic energy of all the cluster's particles with the cluster's Center of Mass (CM) energy removed. The total internal energy of the cluster in the CM frame is the sum of the internal kinetic energy of all cluster's atoms and the potential energy of their interaction. The data on the time-dependent relaxation of the Ar$_n$H$^+$ internal energies, i.e. specifically the kinetic, potential, and total energies of the atoms inside the cluster, have been recorded and analyzed. The relaxation of the cluster's potential energy, which is an essential part of the internal energy, reflects the continuous transformation of the cluster geometrical configuration through the entire cooling process. To establish a time-dependent picture of the relaxation of the internal kinetic and total energies of Ar$_n$H$^+$ clusters, these energies have been computed every ps that a cluster exists and tracked through time along with the time dependent cluster size $n=n(t)$. To illustrate the tracking procedure, the evolution of the cluster internal kinetic energy, denoted as $\varepsilon_{k}$, for an Ar$_n$H$^+$ cluster in a single simulation trajectory is shown in Fig. \ref{fig:SingleClusterKE}.

The results of the single proton simulations depicted in Fig. \ref{fig:SingleClusterKE} also illustrate the time-evolution of the size $n(t)$ of Ar$_{n(t)}$H$^+$ clusters. The dynamics of cluster growth includes both processes, the size increases $n\rightarrow n+1$ and decays $n\rightarrow n-1$, which occur with different probabilities during the non-equilibrium nucleation of small clusters. The probability of $n\rightarrow n+1$ transitions, i.e. the probability of the sticking collisions between clusters and free Ar atoms surpasses the decay probability during the non-equilibrium stage of nucleation. The probabilities of both these processes became equal when the system has reached equilibrium between the gas and cluster phase.

There are several noteworthy observations that can be made analyzing the  cluster energy relaxation. First, we notice the remarkable stability of the Ar$_4$H$^+$ cluster, which has the first closed shell. This cluster generally does not  demonstrate $n\rightarrow n-1$ transitions, as it is extremely stable with deep binding energies of the cluster Ar atoms \cite{OurArticle}. Moreover, the Ar$_4$H$^+$ cluster exists in the broad time-domain  that overlaps significantly with time-regions that belong to other clusters. It means, that under non-equilibrium conditions any decay of larger clusters, such Ar$_5$H$^+$ or Ar$_6$H$^+$, could end up returning with a high probability to the stable configuration of Ar$_4$H$^+$. Second, the growth of small clusters through the collision reaction Ar$_{n-1}$H$^{+}$ + Ar $\rightarrow $ Ar$_{n}^{\star}$H$^{+}$, leads to a sharp increase in the cluster's internal energy. The average value of the release of internal kinetic energy approximately corresponds to the absolute value of the chemical potential of Ar atoms in the Ar$_n$H$^{+}$ cluster\cite{OurArticle}. For small-sized clusters, which can be considered as molecular ions Ar$_n$H$^+$, the $n \rightarrow n+1$ growth leads to multiple excitations of vibrational and rotational modes. Although the growth of small-$n$ clusters causes a significant excitation of the cluster's internal degrees of freedom, the magnitude of this excitation decreases with cluster size $n$ \cite{OurArticle}. The fact that the magnitude of excitation decreases with cluster size $n$ also implies that the cluster becomes progressively easier to grow from an energetic standpoint because there is less internal energy that should be cooled down before this cluster can grow. Additionally, the internal kinetic energy released in $n \rightarrow n+1$ transitions is much larger than $k_BT$, the scale of thermal energy limiting the cluster growth under equilibrium conditions. This means that the excited state of nascent clusters can persist for a relatively long time, which is required for cooling of nascent clusters. This time depends on the size of the cluster and the efficiency of cooling collisions between  Ar$_{n}^{\star}$H$^+$ and cold atoms of the ambient Ar gas. Another notable observation is that the Ar$_{n-1}$H$^+$ + Ar $\rightarrow$ Ar$^*_{n}$H$^+$ sticking collisions can lead to  formation of long-leaving meta-stable clusters. They can dissociate in spontaneous $n \rightarrow n-1$ transitions or could be a channel of fast cooling of the cluster internal energy in the quenching collisions Ar$^*_{n}$H$^+$ + Ar $\rightarrow$ Ar$_{n}$H$^+$ +Ar*. It appears that the time needed for nascent clusters to relax from the initial rotational-vibrational excited state (i.e. the lifetime of the excited Ar$^*_{n}$H$^+$ clusters) plays a role in mediating the growth of Ar$_n$H$^+$. For specific densities and temperatures of the atomic gas, the relaxation of the internal energy acts as a "bottleneck" in the growth process \footnote{There is an analogy between the cluster nucleation and the processes of electron- highly charged ion recombination in the plasma physics, which involves the capture of electrons into highly excited Rydberg states and subsequent relaxation of the ion internal energy \cite{LLkinetics}.}.

 Highly excited states of cluster particles can be successfully described by semi-classical methods of classical molecular dynamics. Despite their small size, Ar$_n$H$^+$ clusters are known to have a significant number of soft-vibrational modes \cite{Ar2_3_structure,Ritschel2007,Ar_n_H+_QM}, which provide a large density of excited states and contribute to the ongoing semi-classical  dynamics of the internal energy relaxation. As shown in Fig. \ref{fig:SingleClusterKE}, the initial formation of small clusters is accompanied by large energy excitation of the internal degrees of freedom, which justifies the use of the semi-classical MD simulations of the nucleation process. 
 
 \section{Ensemble Average Methods}
In order to investigate the stochastic processes that governs the energy relaxation of a cluster, we generated an ensemble of MD trajectories. These trajectories provide insight into the average behavior of the system and allow us to examine the nucleation of independent clusters induced by an individual H$^+$ ion. To further understand the time-dependent kinetics of Ar$_n$H$^+$ nucleation, we calculated ensemble average quantities for clusters containing the same number of Ar atoms $n$, accounting for the time since the creation of Ar$_n$H$^+$ clusters. We conducted 500 independent MD simulations of the cluster growth initiated by a single H$^+$ ion in a metastable Ar bath gas. The resulting MD trajectories represent an ensemble of independently growing Ar$_n$H$^+$ clusters.

 In the present study, several computational steps were implemented to analyze the data obtained from MD simulations of the nucleation of Ar$_n$H$^+$ clusters. Specifically, for each MD trajectory, the data on the formation and evolution of the clusters was analyzed, and any outlying data was removed using a simple clustering algorithm based on temporal information. The purpose of this procedure was to identify the average time interval during which the cluster size remained stable at $n$ rather than being temporarily changed due to close encounters between the Ar$_n$H$^+$ cluster and free Ar atoms. These encounters can lead to a temporary "fictitious growth" from $n$ to $n+1$ when using the DBSCAN algorithm on LAMMPS trajectories to classify clusters based on the geometric proximity of atoms. However, this growth is typically short-lived and returns to the stable size within 1-2 ps. We employed effectively the time-filtering procedure that removed the majority of these temporary "fictitious growth" events from the ensemble data. 

 To study the time-dependent cooling process  in the ensemble of clusters of a specific size $n$, the ensemble data on nascent Ar$_n$H$^+$ clusters were scaled to artificially start at $t=0$. This new effective time of cluster creation at $t=0$ coincides with the first timestep at which the cluster is observed to be the desired size $n$. This scaling is necessary because the moment when an individual Ar$_n$H$^+$ cluster becomes the desired size is random in absolute time. However, by scaling all clusters of size $n$ to start at a common $t=0$, their energy relaxation dynamics can be compared and $t=0$ can be interpreted as the ``time since formation", $t_f$.

 After the scaling described above, all cluster data for the specific size $n$ have a common time domain. This allows to calculate the ensemble average  values at every time since formation with the adopted time step of 1 ps. To calculate the ensemble average of various cluster parameters, such as the cluster internal energy, center of mass energy, internal kinetic energy, potential energy, and the energy distribution functions for the cluster's Ar and H$^+$ particles. These quantities are binned to create time-dependent empirical probability distributions. These distributions are used to calculate the appropriate time-dependent average values at every subsequent point in time after $t=0$. The empirical probability distribution must be rebuilt at every time step after the time since formation due to the ongoing growth and decay of the cluster size $n$. The time evolution of the number of clusters of a given size $n$ will be discussed in the section \ref{sec::Lifetimes}.
 
 \section{Energy Relaxation in the Ensemble}
\begin{figure}[ht]
    \includegraphics[scale=0.266]{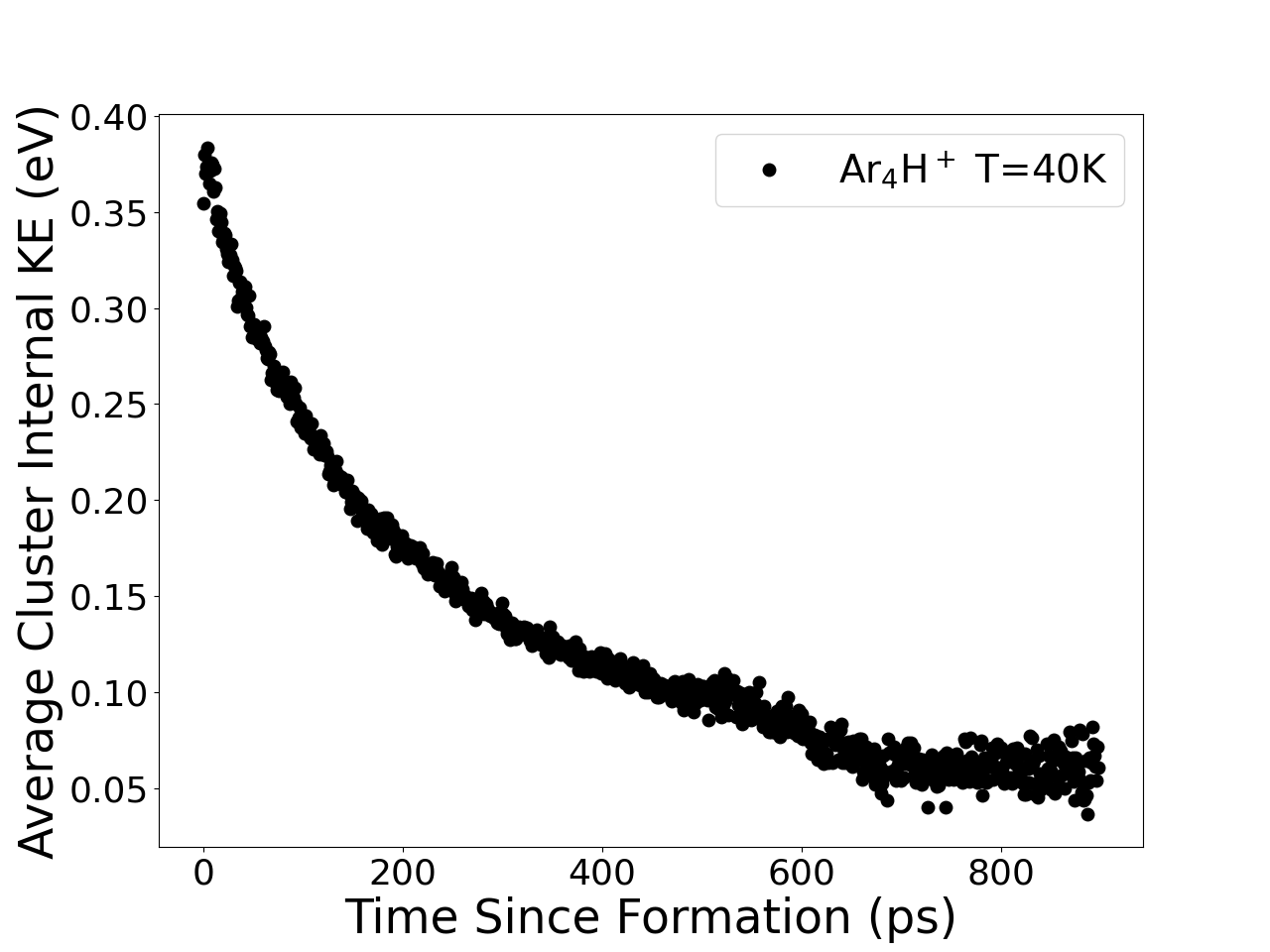}
    \caption{The ensemble average internal kinetic energy $\overline{\varepsilon_{k}}$ is shown for Ar$_4$H$^+$ clusters at a temperature of 40K. The average is calculated from a series of simulations and shows a smooth decrease in internal energy with small fluctuation.}
    \label{fig::InternalKEAr4H}
\end{figure}

\begin{figure}[ht]
    \includegraphics[scale=0.266]{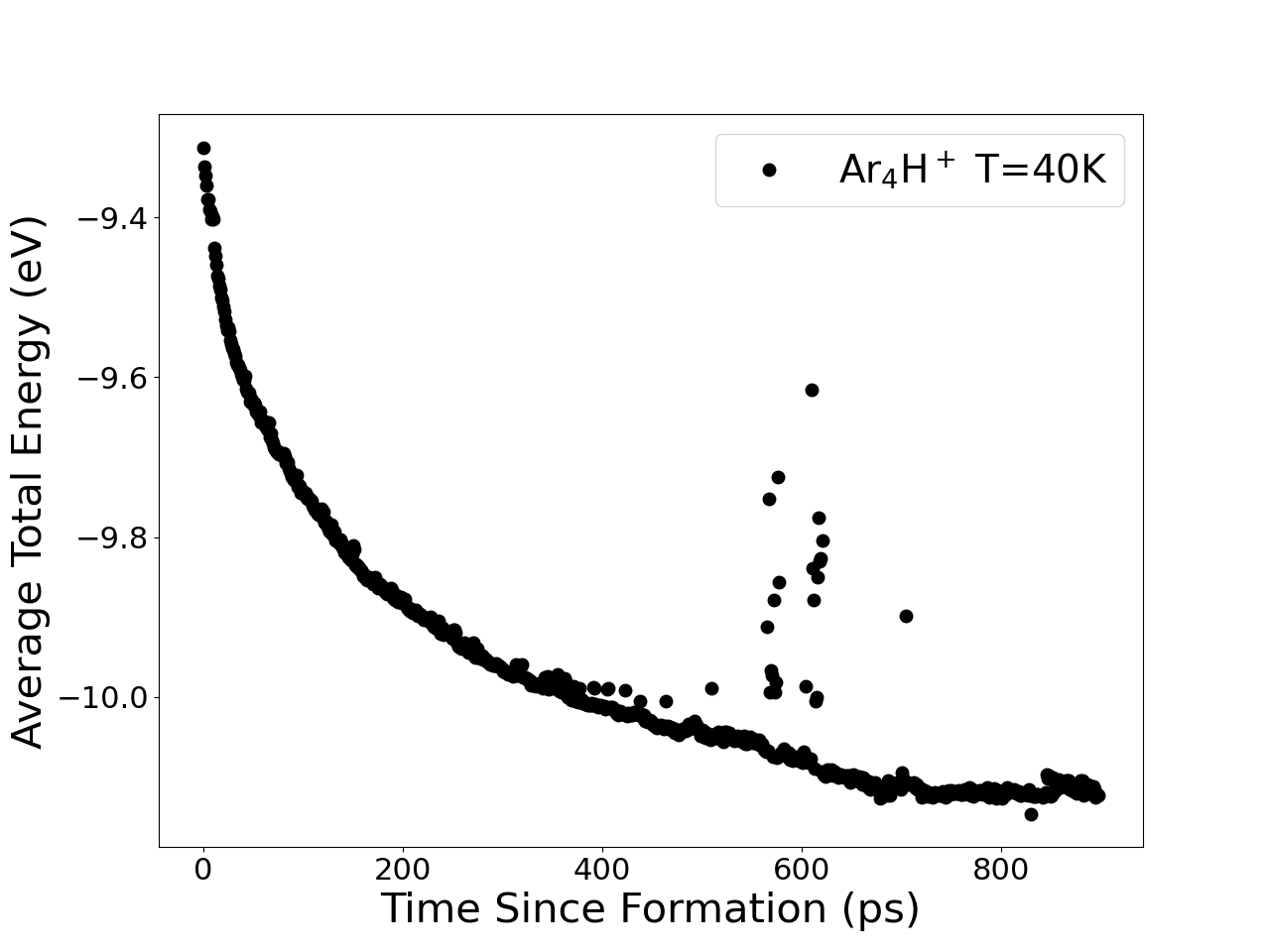}
    \caption{The ensemble average of the total energy of cluster particles $\overline{\varepsilon_{tot}}$ is shown for Ar$_4$H$^+$ clusters at a temperature T=40K. This average is calculated from a series of simulations and shows a smooth behavior except isolated fluctuation near 600ps.}
    \label{fig::TotalEnergyAr4H}
\end{figure}

\begin{figure}[ht]
    \includegraphics[scale=0.266]{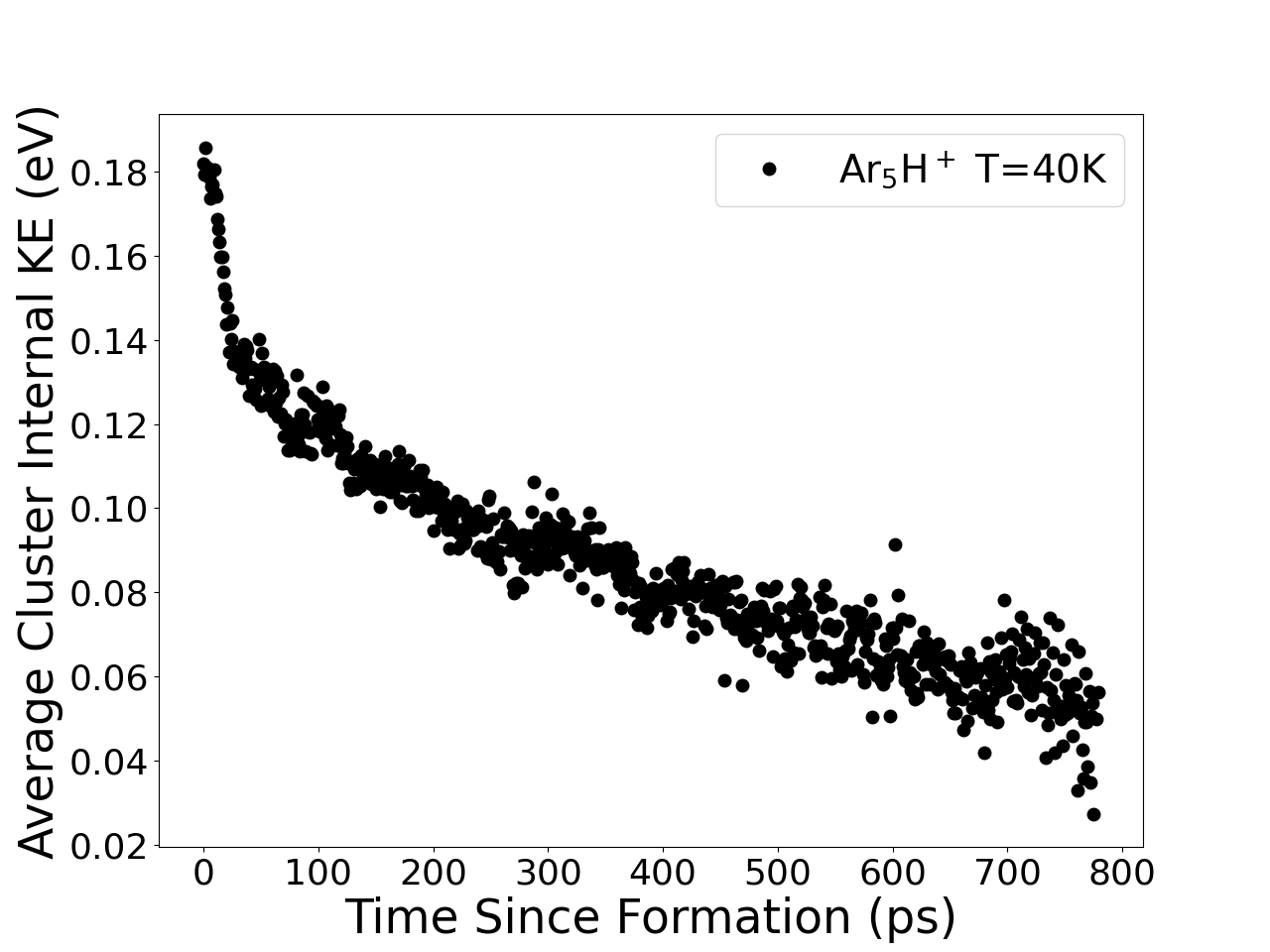}
    \caption{The ensemble average  $\overline{\varepsilon_{k}}$ is shown for Ar$_5$H$^+$ at a temperature T=40K. This average is calculated from a series of simulations and shows a sharp energy relaxation within the first 100 ps, followed by a smooth behavior thereafter.}
    \label{fig::InternalKEAr5H}
\end{figure}

\begin{figure}[ht]
    \includegraphics[scale=0.266]{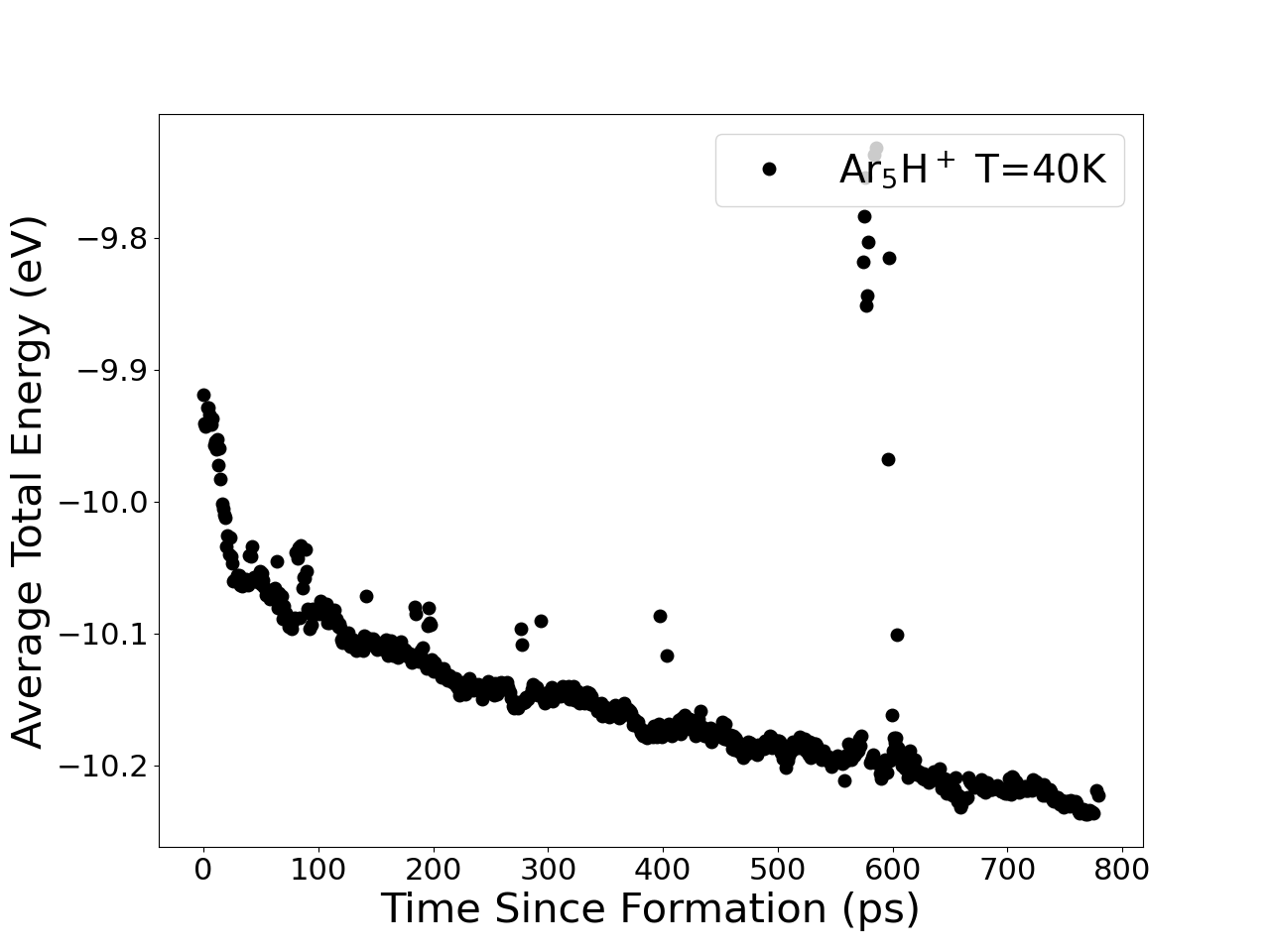}
    \caption{The ensemble average $\overline{\varepsilon_{tot}}$ is shown for Ar$_5$H$^+$ clusters at a temperature T= 40K. The average is based on the results of multiple simulations. A significant increase in the total internal energy, shown as a spike around t=600ps, is observed. This spike is only $\sim$4\% different relative to the surrounding curve. This is attributed to a small group (approximately 20 \% ) of clusters, which initial configurations are very far from the ground state configuration providing cluster's minimal energy. These clusters relax slowly and can contribute to the deviation from the  average total energy.}
    \label{fig::Ar Total E Ave Ar5H}
\end{figure}

The formation of Ar$_n$H$^+$ clusters is initially characterized by non-equilibrium configurations, which can be alternatively represented as a superposition of  highly excited rotational-vibrational states. Through simulations, we have identified the time-dependent kinetics of cluster growth and observed the mechanisms by which the internal kinetic energy of the cluster relaxes. The time-dependent picture of the internal energy relaxation in the ensemble of nascent  Ar$_n^*$H$^+$ clusters is shown through the ensemble average of the internal kinetic, $\overline{\varepsilon_{k}}$ in Figs. \ref{fig::InternalKEAr4H} and \ref{fig::InternalKEAr5H}; and total energies, $\overline{\varepsilon_{tot}}$ in Figs. \ref{fig::TotalEnergyAr4H} and \ref{fig::Ar Total E Ave Ar5H}, for clusters with n=4 and 5 at  T=40K.

In the time-dependence of the internal kinetic energies $\overline{\varepsilon_{k}}$ of clusters with $n=4$ and 5, shown in Figs. \ref{fig::InternalKEAr4H} and \ref{fig::InternalKEAr5H}, we see some notable features. One is that the relaxation is generally smooth, with relatively small fluctuations. The relaxation of the total energy $\overline{\varepsilon_{tot}}$, Figs. \ref{fig::TotalEnergyAr4H} and \ref{fig::Ar Total E Ave Ar5H} is even smoother with smaller fluctuations despite it including the simultaneous fluctuating kinetic and potential (configuration) energies. In our simulations, these clusters never reach the thermal equilibrium. Although their internal energy relaxes during the entire life-time but the cluster kinetic energy is still larger than the characteristic thermal energy of $k_BT$ ($T$=40K). This means that these small clusters have been produced stably and disappeared from their Ar$_n$H$^+$ ensemble due to the size growth. Moreover, they reduced only a fraction of their internal energy, and not relaxed all the way to the thermal energy before they can grow in $n\rightarrow n+1$ transition. 

One of the objectives of this study is to investigate the relationship between the average kinetic energy of cluster particles per internal degree of freedom $\overline{\epsilon_n}$ and the kinetic energy of their center of mass (CM). For all values of $n$ examined, the CM kinetic energy was found to be significantly lower compared to $\overline{\epsilon_n}$. However, as the internal degrees of freedom of the cluster approached thermalization, both the CM kinetic energy and $\overline{\epsilon_n}$ became comparable to the scaling thermal energy $k_BT$. Our results suggest that studying the relaxation of internal energy is an effective approach for characterizing the dominant processes occurring within the cluster.

The internal kinetic energy of clusters exhibits fluctuations that tend to increase in proportion to the cluster size and the time since formation. This is the case both in the deviation from the time average of $\overline{\varepsilon_{k}}$ and in the standard deviation of the mean $s_m$ for each data point. For example, in Fig.\ref{fig::InternalKEAr4H}, we find $s_m(t_f=100ps)$=4.8 meV, $s_m(t_f=500ps)$=7.7 meV, and $s_m(t_f=700ps)$=12 meV. This corresponds to the relative error increasing from $\sim$2\%, to $\sim$7\%, and to $\sim$17\%. This effect is demonstrated in Figs.\ref{fig::InternalKEAr4H} and \ref{fig::InternalKEAr5H},  where the ensemble average tends to broaden as time increases, this broadening is described by increase of the $s_m$ magnitude with the relaxation time. The relative magnitude of energy fluctuations generally becomes more significant in larger clusters.

There are two primary factors contributing to the observed fluctuations in the energy of clusters with larger sizes (n=5 and above). The first factor is lower $\overline{\varepsilon_{k}}$ of the atoms in larger clusters compared to smaller clusters ($n \leq 4$). The second factor causing the time-dependent increase in fluctuations is purely statistical and arises from the depopulation of the ensemble of clusters with a given n due to non-equilibrium cluster growth $n \rightarrow n+1$. Fluctuations of the total cluster energy are relatively small compared with fluctuations of their kinetic energy.

Another notable feature can be seen in Figs.\ref{fig::TotalEnergyAr4H} and \ref{fig::Ar Total E Ave Ar5H}, which are the sharp short-term fluctuations in the total energy in the late stage of the relaxation. These fluctuations appear large in absolute values but have a relative magnitude of only $\sim$4-5\% from the surrounding smooth curves, and look large due to the figure scales. These specific fluctuations don't exist in the kinetic energy time-curves and  arise due to fluctuation of the potential energy of clusters, i. e. fluctuations of their geometrical configurations. Some of the clusters present in the ensemble at the time of these sharp fluctuations have initial configurations that are far from the equilibrium ones discussed in \cite{OurArticle}. These nascent clusters are highly excited and the rates of their energy relaxation and growth differ from the rest of the average relaxation curve. As previously mentioned, the number of clusters in the Ar$_n$H$^+$ ensemble decreases with the time-interval since formation and the ensemble average values become more sensitive to the contribution of clusters with initial configurations which are very far from equilibrium ones. This means that the ensemble-averaged values at a later time are more sensitive to the contribution of clusters with initially highly excited potential energies. 

\begin{figure}[ht]
    \centering
    \includegraphics[scale=0.266]{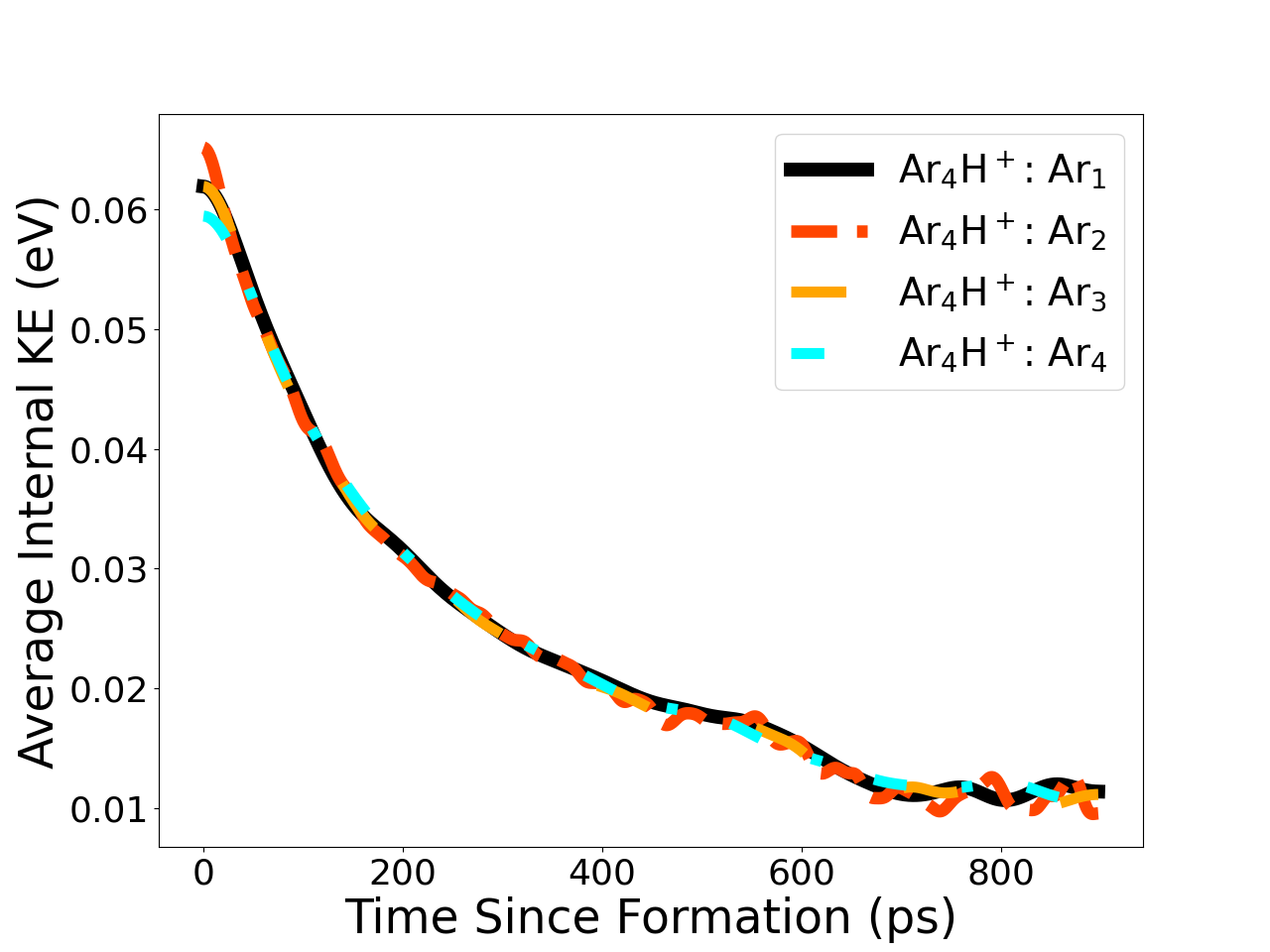}
    \caption{Shown is $\overline{\epsilon_{k,i}}$ for the individual Ar atoms annotated as Ar$_i$ with i=1-4inside a Ar$_4$H$^+$ cluster. The average internal kinetic energy data has been smoothed to compare the time averaged behaviour of the individual Ar atoms. This average is calculated from the ensemble of simulations at T=40K. Although there are fluctuations present (which are dampened by the smoothing), the behavior of each individual Ar atom is identical on average in time. The solid line represents Ar$_1$, the dashed-dot line represents Ar$_2$, the dashed line represents Ar$_3$, and the large spaced dashed-dot line represents Ar$_4$.}
    \label{fig:ArComparisonAr4}
\end{figure}

In order to compare the internal kinetic energies $\epsilon_{k,Ar(i)}$ (where i=1-4) of different Ar atoms within a cluster and the energy of the H$^+$ ion, $\epsilon_{k,H^+}$ the energies of the cluster's individual particles were calculated. We first performed an ensemble average on the internal kinetic energies of individual Ar atoms bound inside an Ar$_n$H$^+$ cluster. We designated different Ar atoms as "i=1", "i=2," and so on based on their position in an array generated by the DBSCAN algorithm, without considering their specific locations within the cluster. This allowed us to consistently track the Ar atoms across different clusters of the same size within the ensemble. The simulation results for the ensemble average of the internal kinetic energies, $\overline{\epsilon_{k,i}}$, of four different Ar$_i$ atoms ($i=1-4$) within the Ar$_4$H$^+$ cluster are shown in Fig.\ref{fig:ArComparisonAr4} at a temperature of T=40K. 
In Fig.\ref{fig:ArComparisonAr4}, the kinetic energy of individual Ar atoms within the Ar$_4$H$^+$ cluster exhibits a similar time dependence, indicating that the kinetic energy of different Ar atoms within the cluster quickly becomes equalized. The smoothed fluctuations in the kinetic energy of each Ar atom, depicted in Fig.\ref{fig:ArComparisonAr4}, are relatively small compared to the actual energy. For larger values of $n$ (i.e., $n \geq 5$), the fluctuations relative to the actual energy values increase due to the reduced excitation of individual Ar atoms in clusters with large $n$, but the kinetic energy equalization of Ar atoms inside clusters is still much faster than other relaxation processes. Therefore, the data for various Ar atoms in a cluster of a specific size $n$ can be combined to generate the empirical energy distribution function for a single Ar atom inside the Ar$_n$H$^+$ cluster at a particular time-step. Moreover, in large clusters, the potential and total energies will not be equal for Ar atoms from different shells, due to the large difference in the strong interaction between the H$^+$ and first shell Ar atoms; while the kinetic energy will equalize between the cluster's atoms. This allows for greater flexibility in modeling the relaxation and growth processes using the empirical single-particle distribution function.

\begin{figure}[ht]
    \includegraphics[scale=0.266]{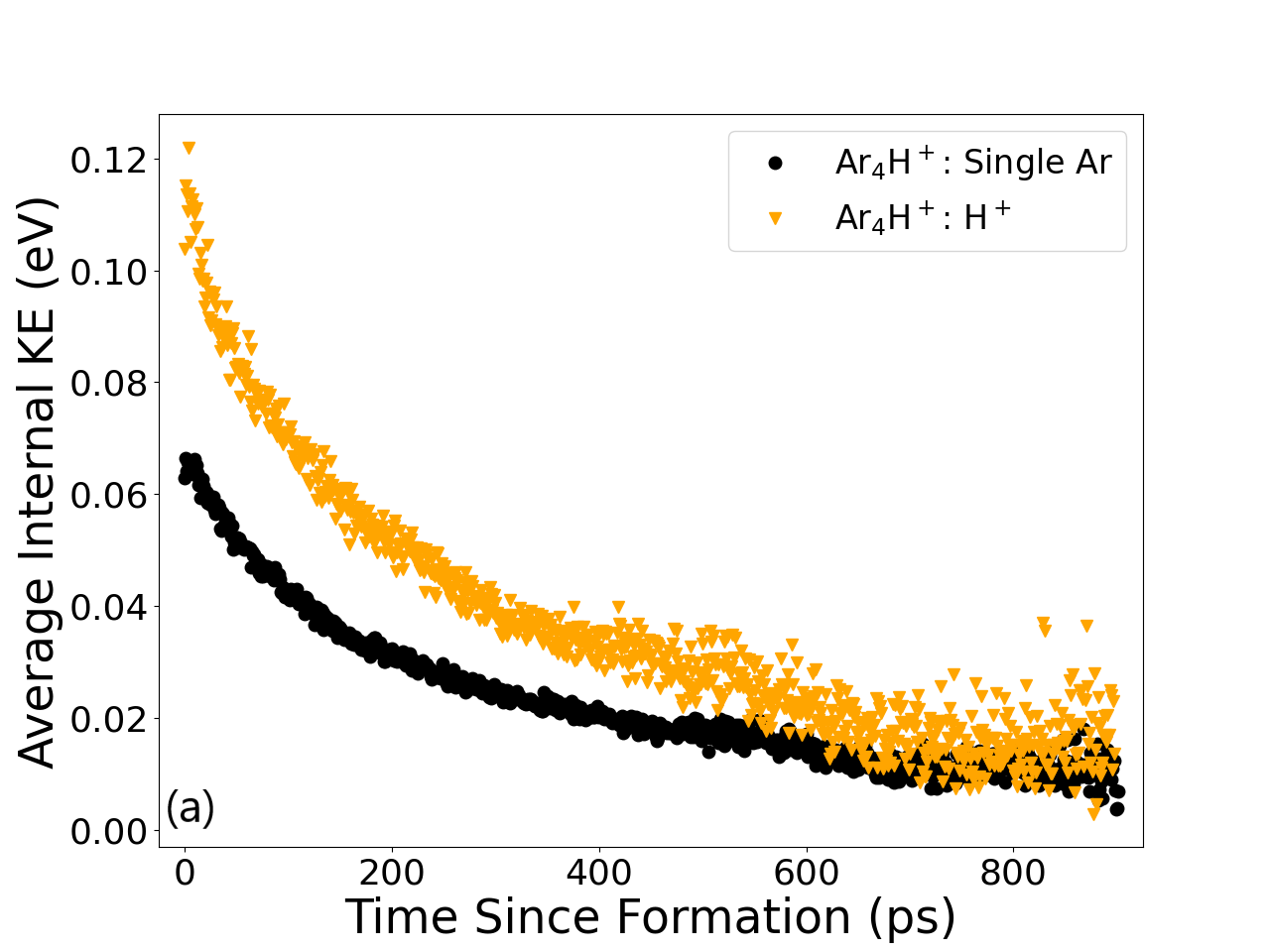}
    \includegraphics[scale=0.266]{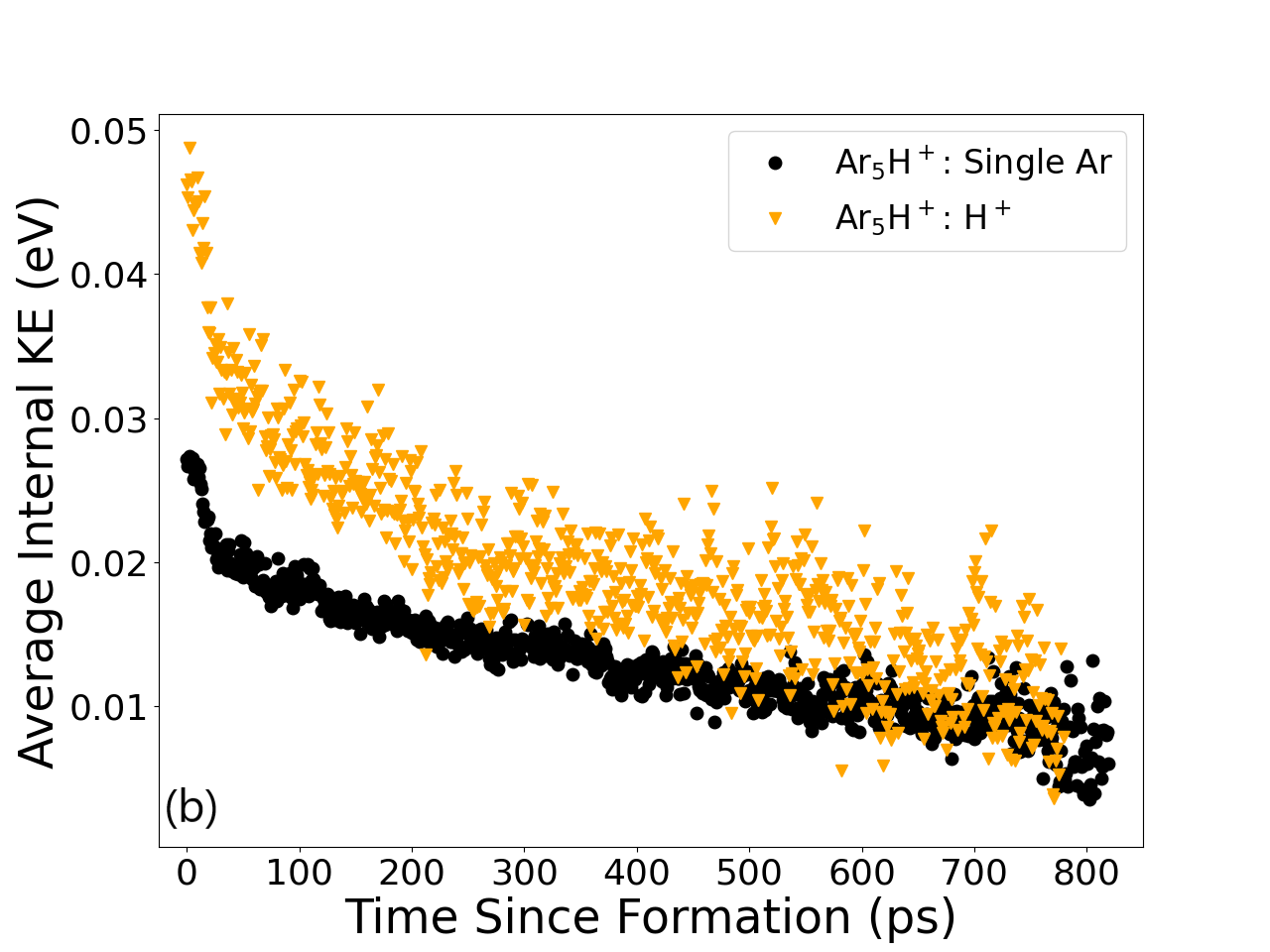}
    \caption{Shown is $\overline{\epsilon_{k,Ar}}$ and $\overline{\epsilon_{k,H^+}}$ inside a Ar$_4$H$^+$ (a) and Ar$_5$H$^+$ (b) cluster. This average is calculated from the ensemble of simulations at T=40K. It is notable that the H$^+$ ion has $\sim$ 2 times the energy of a single Ar atom during the clusters' initial formation.}
    \label{fig:HArComparison}
\end{figure}

The other comparison to be made is between the energies of a randomly chosen Ar atom and the H$^+$ ion inside the Ar$_n$H$^+$ cluster. The ensemble-averaged internal kinetic energies for a cluster's single randomly chosen Ar, $\overline{\epsilon_{k,Ar}}$, and H$^+$, $\overline{\epsilon_{k,H^+}}$, particles have been inferred from the results of simulations for the entire relaxation process using the previously described methods. The results are shown for Ar$_n$H$^+$ clusters with n= 4 and 5 at T=40K in Figs. \ref{fig:HArComparison}a and \ref{fig:HArComparison}b.

Several comments can be made based on these results. For the entire relaxation process in the small Ar$_n$H$^+$ clusters, the H$^+$ ions have more kinetic energy than a single Ar atom. As the cluster size $n$ increases, the difference in kinetic energies between the H$^+$ ions and Ar atoms decreases with $n$ as it is shown in Figs.\ref{fig:HArComparison}a and  \ref{fig:HArComparison}b. Eventually, this gap in kinetic energies disappears as the entire cluster approaches the thermal equilibrium. In the energy relaxation process, the difference in kinetic energies of the cluster's H$^+$ ion and Ar atoms is expected due to the large mass difference between these two particles. The lighter H$^+$ ion has  the slower rate of energy transfer and thus the slower thermalization compared to heavy Ar atoms. This mass difference explains the more pronounced fluctuations in the kinetic energy of the H$^+$ ions. There is also a strictly statistical reason for the reduction of the energy fluctuations of Ar atoms compared to those of the H$^+$ ion. Since the ensemble average for a single Ar atom is done over both the ensemble of simulations and all Ar atoms in the Ar$_n$H$^+$ cluster, there is a factor of $n$ more data points for the single Ar results compared to the H$^+$ ion data.

\section{\texorpdfstring{Non-equilibrium Growth and Lifetimes of $Ar_nH^+$ Clusters}{Section:Lifetimes}}\label{sec::Lifetimes}
\begin{figure}[ht]
    \includegraphics[scale=.266]{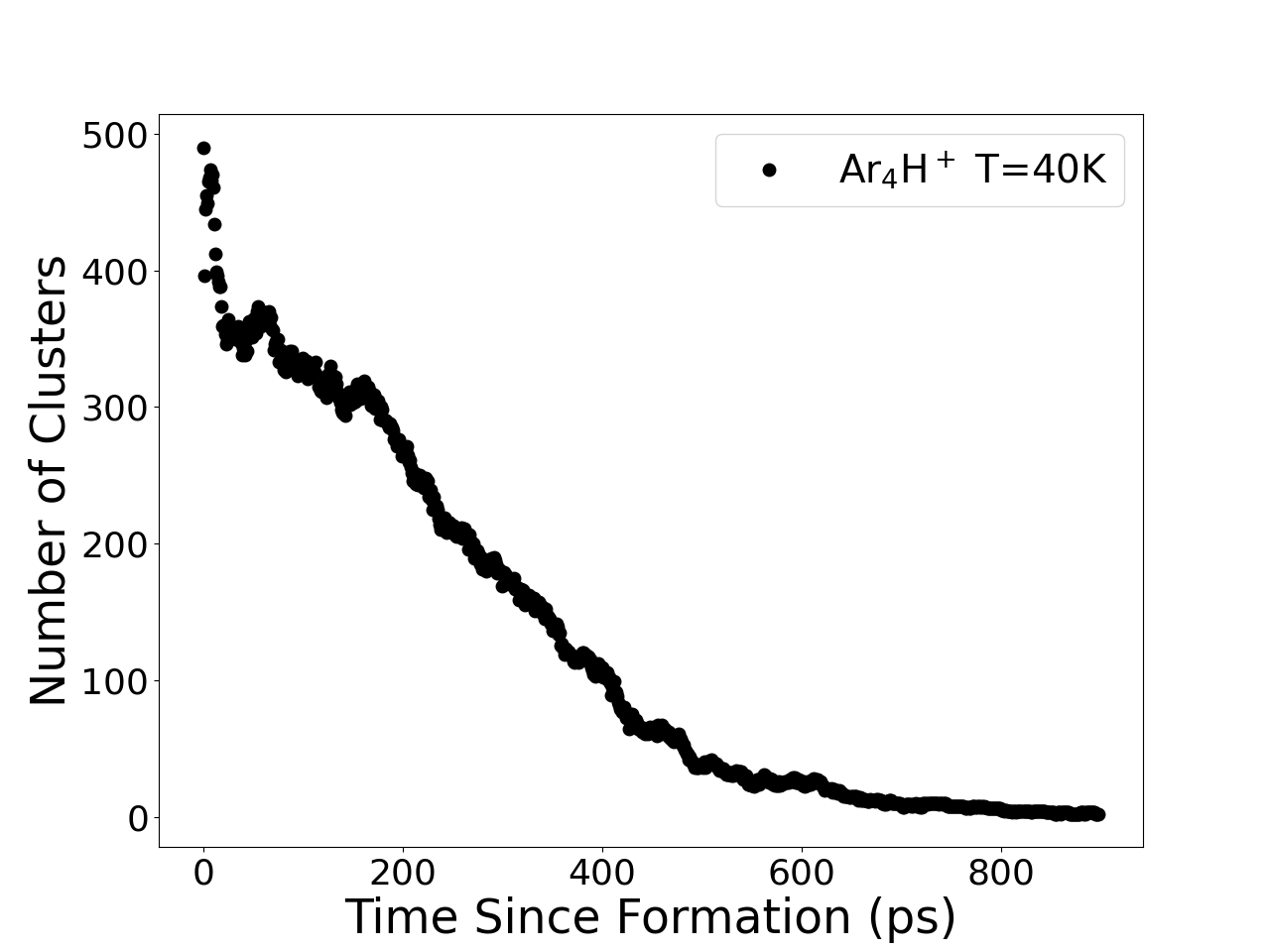}
    \caption{Shown is the number of Ar$_4$H$^+$ clusters present in the ensemble since the time of formation. The number of clusters is inferred from the ensemble of simulations at T=40K. We see that the decrease is relatively smooth across all times since formation.}
    \label{fig:NumPointsAr4H}
\end{figure}

\begin{figure}[ht]
    \includegraphics[scale=.266]{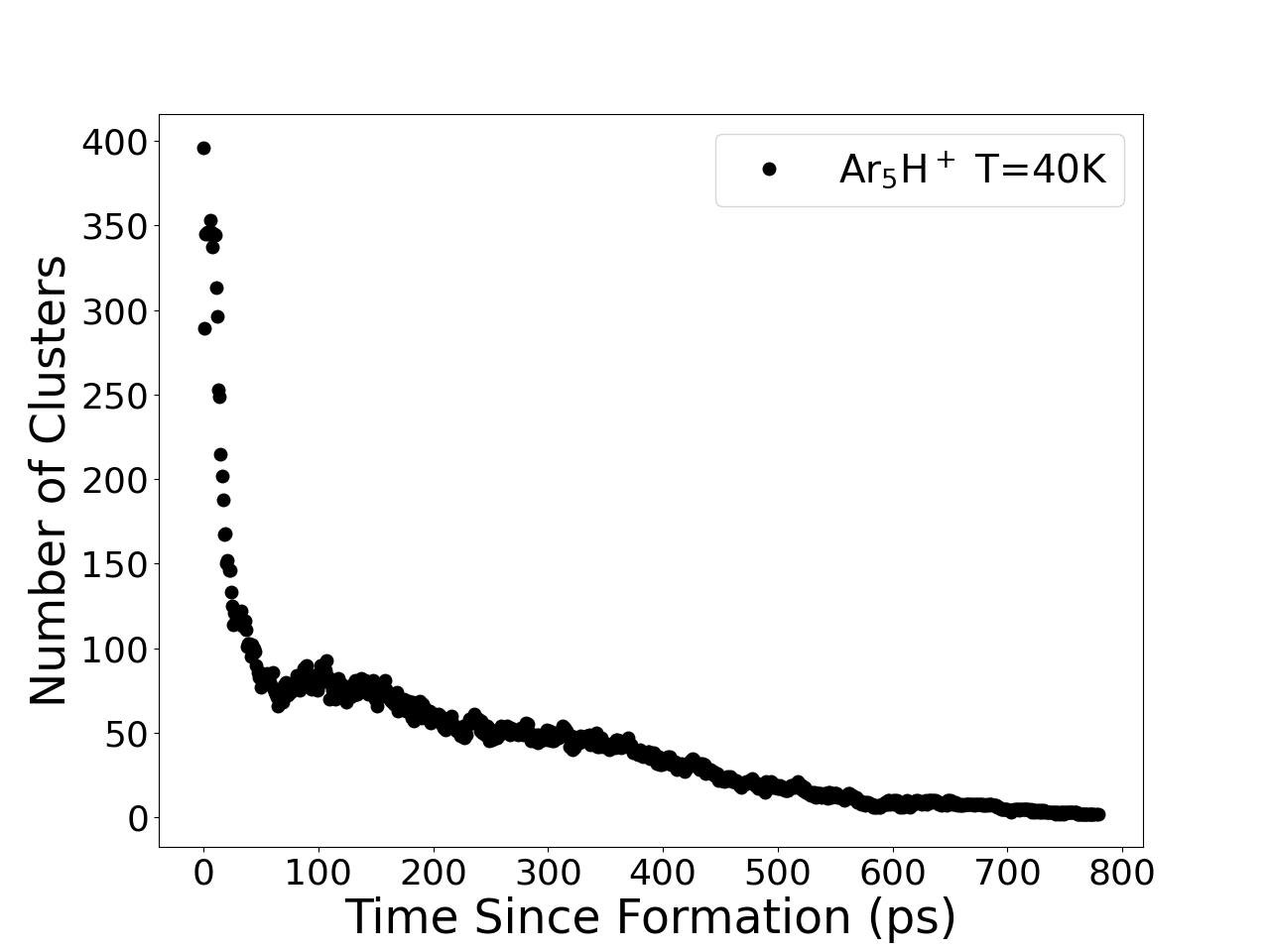}
    \caption{Shown is the number of Ar$_5$H$^+$ clusters present in the ensemble since the time of formation. The number of clusters is inferred from the ensemble of simulations at T=40K. We see that there is a sharp decrease in the first $\sim$ 60 ps followed by relatively flat behaviour and a smooth decrease.}
    \label{fig:NumPointsAr5H}
\end{figure}

In our simulation ensemble, we tracked the time-dependent velocities and coordinates of all particles inside the clusters of size $n$ and also recorded the number $N(n,t)$ of such clusters as a function of the time since their formation. This allowed us to observe a notable behavior within the ensemble of growing nano-particles, specifically the time-dependent growth and depletion of the number of the Ar$_n$H$^+$ clusters of the selected size $n$. The time dependencies of the number of clusters $N(n,t)$ with $n \leq 11$ have been inferred from the database on the non-equilibrium cluster growth in the ensemble of 500 independent clusters. To exemplify typical features of the time-evolution of the cluster population $N(n,t)$, we show the inferred number of Ar$_4$H$^+$ and Ar$_5$H$^+$ clusters as the functions of time since their formation in Figs. \ref{fig:NumPointsAr4H} and \ref{fig:NumPointsAr5H}. 


The depicted small fluctuations in the number of clusters reflect the stochastic nature of the nucleation process. Two types of behaviors of the ensemble population $N(n,t)$ are observed. For very small clusters with $n \leq 4$, the depopulation curves $N(n,t)$ are relatively smooth, as illustrated in Fig. \ref{fig:NumPointsAr4H} for n=4, until the majority of the clusters in the initial ensemble is cooled down and undergo a size-change due to $n \rightarrow n+1$ transitions. The relatively fast but small initial drop of the $N(n,t)$ values, as it is shown in Fig.\ref{fig:NumPointsAr4H}, can be explain by the small fractions of the meta-stable clusters decaying in $n\rightarrow n-1$ transitions and $n$-size growth of 
 Ar$_n$H$^+$ clusters with highly exited configurations. The latter cluster can quickly leave Ar$_n$H$^+$ ensemble due to efficient long-range sticking $n\rightarrow n+1$  or dissociative $n\rightarrow n-1$ collisions with the cold bath gas atoms. 

In the protonated Ar clusters with $n\leq 4$,  all atoms belonging to the first, most stable and tight atomic shells \cite{OurArticle}. The growth of Ar$_n$H$^+$ with $n \geq 4$ leads to the formation of the outer, loosely bound solvation shells. The presence of inner and outer atomic shells  determines the distinct time-behavior of the non-equilibrium population $N(n,t)$. 
Simulations show a sharp decrease in the number of clusters in the Ar$_n$H$^+$ ($n\geq 5$) ensembles  within the first $\sim$ 50-80 ps, followed by a slower stage of  $N(n,t)$ decrease that takes much longer compared to the initial rapid decrease. The example of this general structure is shown in Fig. \ref{fig:NumPointsAr5H} for the nascent Ar$_5$H$^+$ clusters. Their depopulation is characterized by two distinct time scales: one that is fast, around 60 ps, and another that is much longer, approximately 700-800 ps. The rapid phase of depopulation coincides with a sharp drop ($\sim 15-20\%$) in the value of average internal kinetic energy, as shown in Fig. \ref{fig:HArComparison}b. This correlation will be discussed in the next section.

Under conditions of the non-equilibrium growth, the life-time $\tau_n$ of Ar$_n$H$^+$ clusters can be defined as an ensemble-averaged time-interval required for the decay of the initial non-equilibrium population of Ar$_n$H$^+$ with the following steady growth of Ar$_{n+1}$H$^+$. The degradation of the cluster population $N(n,t)$ occurs in multiple $n\rightarrow n\pm1 $ transitions in cluster collisions with Ar atoms of the ambient gas. In our simulations, the non-equilibrium rate of cluster growth in $n\rightarrow n+1$ collisions is larger than the rate of the detachment processes $n\rightarrow n-1$. The smaller clusters with only inner atomic shell $n \leq 4$, have longer lifetimes compared to ones with $n \geq 5$. This behavior is completely consistent with the general statement that probability of collisional capture of Ar atoms from the cold bath gas increases significantly with the reduction of the energy of cluster particles belonging to the outer cluster shells. The magnitude of the excitation on nascent clusters in $n\rightarrow n+1$ process is sharply decreasing with $n$, and becomes roughly constant for $n \geq 5$. This is a reflection of $n$-dependence of the binding energy of Ar atoms in the outer shells of Ar$_n$H$^+$ clusters.
 
Several collision processes contribute to the fast cluster growth and so to the depopulation of  Ar$_n$H$^+$ ($n \geq 5$) ensembles of the nascent clusters with the outer shells of Ar atoms. The nascent Ar$_n^*$H$^+$ ($n\geq 5)$ clusters are formed in highly excited states with a large variety of excited configurations provided by outer shell Ar atoms. These atoms significantly impact the rates of cluster energy relaxation and growth. The simplified model of cluster nucleation dynamics presented below is not exact, but it can be rationalized for clusters with $n\geq 5$, owing to the significant contrast in the binding energies of Ar atoms in the first and second atomic shells \cite{OurArticle}.

The configurations of the nascent excited Ar$_n^*$H$^+$ can be approximately classified into two groups: (a) long-range configurations and (b) short range (tight) configurations. In the (a)-type configurations, incoming Ar atoms are captured in the long-range but shallow attractive part of the potential of interaction between the Ar$_{n-1}$H$^+$ cluster core and the incoming free Ar atom. For example, for the excited  Ar$_5^*$H$^+$, the capture of a free Ar atom into the outer shell with long-range configuration occurs mostly in Ar+Ar collisions of free atoms inside long-range potential field of Ar$_4$H$^+$ cluster core. This is a direct analogy to the 3-body recombination processes in a dense plasma \cite{LLkinetics}, where the cluster core plays the role of the "third" body. As these collisions occur within the shallow region of the potential well, both the kinetic and potential energies of the  captured Ar atoms in the outer shell are comparatively low. The Ar atoms in outer can efficiently exchange energies  with the cold bath gas and, at the same time, participates in cooling of hot atoms of the cluster core. Notice, that for $n \geq 5$ cluster both the cluster core and outer shell can be in excited states.  

Then (b)-type (tight) configurations are mostly formed in direct sticking collisions between the cluster core and free Ar atoms. Here in the cluster core region, the kinetic energy of incoming free Ar atom becomes large and can be transferred to core particles, e.g. in an inelastic collision exciting internal rotational-vibrational states. This will cause the outer shell Ar atom to remain near the core region for the entire relaxation process and the outer shell Ar atom in this "tight" configuration will have relatively large values of the initial  kinetic energy. The (b)-type of configurations are more close to the equilibrium one for than the long-range configurations (a). The energy relaxation of the (b)-type  nascent Ar$_5^*$H$^+$ clusters looks similar to the slow relaxation of the strongly bound and tight Ar$_4^*$H$^+$.

Our simulations were performed at the high density of the Ar bath gas (10$^{20}$cm$^{-3}$). For this very high density the rate of sticking collisions populating (a)-configurations is higher than the rate of direct collisions with the cluster core creating the (b)-type of clusters. The phase volume of the long-range (a)-configurations is significantly larger than that for the tight configurations (b). Thus, we expect that the ensemble population of nascent Ar$_n^*$H$^+$ ($n \geq 5$)clusters with outer shells are mostly formed and decay due to population of the (a)-type long-range configurations. Estimates for the Ar$_5^*$H$^+$ clusters, based on the analysis of simulation results and simplified modeling of Ar$_4$H$^+$ + Ar potential, shown that about 60-70\% of nascent Ar$_5^*$H$^+$ are formed with the long-range configurations (a).
 
There are a variety of collision processes in each configuration, (a) and (b), that can lead to the significant and fast drop in cluster population in Fig. \ref{fig:NumPointsAr5H}. For the (a)-type of clusters  with the outer shell atoms, all growth and relaxation processes  are expected to be relatively fast,$\sim$ 50-80ps, while the "tight" configurations provide the longer timescale of the $n\rightarrow n+1$ growth transitions.

The detailed analysis of any specific type of sticking (inelastic) collisions between free atoms and molecules or  small clusters, such as Ar$_n$H$^+$ +Ar $\rightarrow$ Ar$_{n+1}^*$H$^+$, requires specific quantum-mechanical or semi-classical calculations \cite{VasiliRelaxation,Stoecklin2016,Hansen2014}, which are a part of our future projects. Nevertheless, we can describe  a general feature of the sticking collision between clusters and atomic particles. In isolated nano-size systems with a finite value of the phase space, the small system is expected to return to its initial state after some finite amount of time $\tau_P$, which is known as the Poincaré recurrence time \cite{Poincare}. If $\tau_P$ is much longer than the time between consecutive cluster collisions with the bath gas atoms ($\tau_P$ $\gg$ t$_c$), the excited nascent cluster Ar$_{n+1}\textsuperscript{*}$H$^+$ can reduce internal energy in cooling collisions with the ambient gas and stabilize a $n \rightarrow n+1 $ size-changing transition. However, for very small clusters $\tau_P$ can be shorter than the time required for internal energy relaxation of nascent Ar$_{n+1}^*$H$^+$ clusters and cluster growth will be terminated in $n+1 \rightarrow n$ transitions. Growing in size cluster will gain more internal degrees of freedom. That leads to an increase of the recurrence time $\tau_P$ and rate of the cluster growth. 

\section{\texorpdfstring{Collisional Cooling of Nascent $Ar_nH^+$ Clusters}{Section: Cooling}}

Collisions between bath gas atoms and excited  Ar$_n^*$H$^+$ clusters result in the relaxation of the internal energy of the nascent clusters and initiate their growth. Our MD simulations provide detailed information on the time-dependent relaxation of the cluster internal energies. To analyze the simulation data on the cooling of small clusters we have used the numerical solutions of the Boltzmann kinetic equation. The accurate quantum-mechanical potentials have been used to study the energy transfer in the binary collisions between cluster particles and atoms of the Ar bath gas. Computational methods were developed in our previous study of the time-dependent energy relaxation of hot atomic particles in  molecular and atomic gases \cite{HydrogenRelaxation,NitrogenRelaxation,SXeRelaxation,KHARCHENKOFastnNitrogen}. 

The kinetic energies of atomic particles belonging to the nascent clusters are significantly higher than thermal ones, and the energy relaxation of cluster's particles in collisions with the cold Ar atoms can be described as the cluster cooling. The computational scheme has been significantly simplified because the encounter between cluster's Ar and free Ar atom is sufficiently rapid due to the short range of Ar potentials. Such encounter can be considered as a spatially localized "point" collision, resulting in an exchange in the particle's kinetic energies exclusively. Accurate rates of energy transfer, i.e. rates of cluster cooling, can be calculated using known differential cross sections of elastic and inelastic collisions of atomic or molecular particles \cite{KHARCHENKOFastnNitrogen,SXeRelaxation}. For our purposes, the simplest hard-sphere approximation (HSA) has been utilized as the collision model for the short-range forces associated with Ar atoms \cite{AndersonHardSphere,KHARCHENKOFastnNitrogen}. Although H$^+$ + Ar potential includes a long-range tail arising from the Ar's polarization, the energy transfer in H$^+$ and Ar collisions can be also described by the HSA model with an appropriate transport cross section. We computed the time-dependent kinetic energy distribution function for hot cluster particles using the Boltzmann equation:
\begin{eqnarray}\label{eq:Boltzmann}
    \frac{\partial}{\partial t}f_n(\epsilon_k,t) =&& \int B(\epsilon_k|\epsilon'_k)f_n(\epsilon'_k,t)d\epsilon'_k \nonumber\\
    &&- f_n(\epsilon_k,t)\int B(\epsilon'_k|\epsilon_k)d\epsilon'_k +Q_n(\epsilon_k,t) ,
\end{eqnarray}
where $f_{n}(\epsilon_k,t)$ is the kinetic energy distribution function for the hot Ar or H$^+$ particles within a Ar$_n$H$^+$ cluster, $B(\epsilon_k|\epsilon'_k)$ is the energy relaxation kernel representing the rates of $ \epsilon'_k \rightarrow \epsilon_k $ transitions due to collisions of cluster particles with the bath gas atoms, and $Q_n(\epsilon_k,T)$ is the sink or source function describing the the production or removal of the Ar$_n$H$^+$ clusters with the specific kinetic energies $\epsilon_k$ of the cluster's particles. Under conditions of the strong non-equilibrium growth, the rate of $n\rightarrow n+1$ transitions dominates and this depopulates the initial non-equilibrium Ar$_n$H$^+$ ensemble. In the Appendix, we provide an analytical expression for the $B(\epsilon_k|\epsilon'_k)$ kernel and discuss details of the numerical solution of the Boltzmann equation. 

In the framework of the relaxation-time approximation \cite{LLkinetics}, the sink term $Q_n(\epsilon_k,t)$  in Eq.\ref{eq:Boltzmann} can be expressed via the empirical life-time $\tau_{n}(\epsilon_k)$  of Ar$_{n}$H$^+$ clusters:
 \begin{equation}\label{eq:DecayTerm}
   Q_n(\epsilon_k,t) = -f_n(\epsilon_k,t)/\tau_{n}(\epsilon_k),
\end{equation}
where $\tau_n(\epsilon_k)$ is the characteristic time describing the reduction of Ar$_n^*$H$^+$ cluster populations due to the $n\rightarrow n +1$ growth transitions. The relationship between the time-scales of the cluster cooling and growing processes depends on the number of cluster's Ar atoms $n$, which also defines the number of cluster's atomic shells. All relaxation processes, including the cluster growth and cooling, occur significantly faster in the clusters with the outer shells, as it was discussed in the Section VI.

The cluster cooling and the cluster growth both take place due to collisions with Ar bath gas atoms. In our simulations we observed two main scenarios for the non-equilibrium cluster nucleation. The first scenario happens when the clusters cool down faster than they grow. The second scenario, on the other hand, manifests the opposite situation where the rate of cluster growth is faster than the rate of cooling of the internal degrees of freedom.

Scenario 1. \textit{The relaxation of the cluster internal energy regulates cluster growth}. This "cooling and growth" scenario works for very small clusters with $n \leq 4$, where the equalization of energies of Ar atoms belonging to the deepest atomic shell occurs very quickly due to strong Ar-Ar and Ar-H$^+$ interactions at relatively small distances between atomic particles inside the clusters. Under such conditions, the cooling collisions with the bath gas atoms influence simultaneously all cluster particles. The energy relaxation rates for every Ar atom and their mean energy are about the same for the entire cooling process, as it is shown in Fig.\ref{fig:ArComparisonAr4}. Although the initial kinetic energy of the light cluster particle H$^+$ is larger than  energy of heavy Ar atoms, as shown in Figs.\ref{fig:HArComparison}a and \ref{fig:HArComparison}b, both relaxation processes are accomplished with the same cooling time due to efficient energy exchange between cluster particles. They are strongly correlated and their energy relaxation processes well described by the Boltzmann equations. 

\begin{figure}[ht]
    \includegraphics[scale=.266]{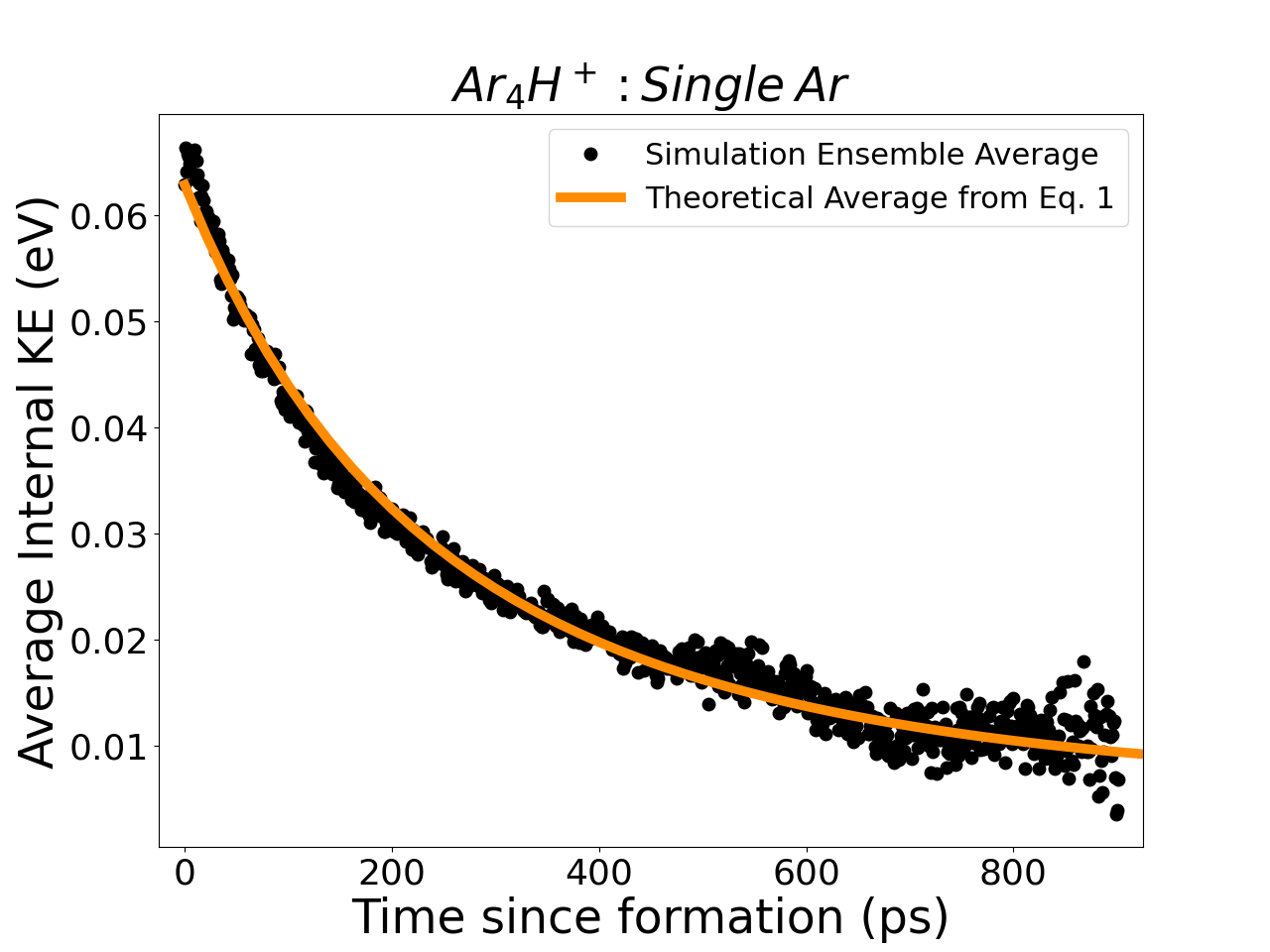}
    \caption{Shown is $\overline{\epsilon_{k,Ar}}$ inside a Ar$_4$H$^+$ cluster, from the ensemble of T=40K simulations, alongside the theoretical time-dependent average calculated from the Boltzmann equation using the initial distribution function found directly from our ensemble of simulations as the initial distribution.}
    \label{fig:Ar fit Ar4H}
\end{figure}

\begin{figure}[ht]
    \includegraphics[scale=.266]{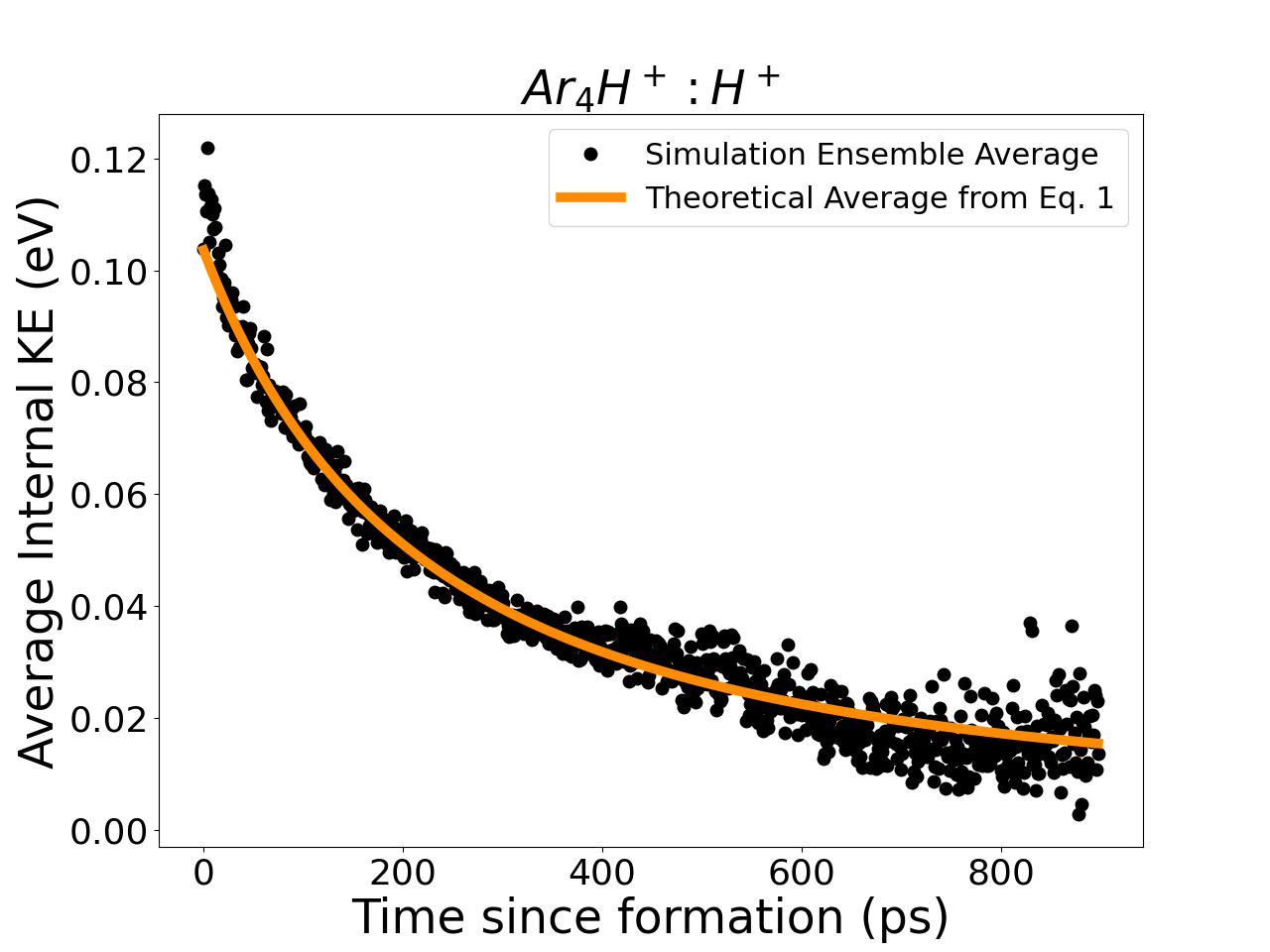}
    \caption{Shown is $\overline{\epsilon_{k,H^+}}$ inside a Ar$_4$H$^+$ cluster, from the ensemble of T=40K simulations, alongside the theoretical time-dependent average calculated from the Boltzmann equation using the initial distribution function found directly from our ensemble of simulations as the initial distribution.}
    \label{fig:H fit Ar4H}
\end{figure}

The time-dependent energy distribution functions, $f_n(\epsilon_k,t)$, for the hot Ar and H$^+$ particles inside nascent Ar$_n$H$^+$ clusters have been obtained through the numerical solutions of the Boltzmann equation with the initial distribution function extracted directly from our ensemble of simulations. The obtained solutions of  Eq.\ref{eq:Boltzmann} were used for computations of the time-dependent average kinetic energies $\overline{\epsilon_{k,Ar}}$ of Ar and  $\overline{\epsilon_{k,H^+}}$ of H$^+$ cluster particles. The results of calculations are in very good agreement with the data inferred from the time simulations of the growth and cooling of small with $n\leq 4$. In Figs. \ref{fig:Ar fit Ar4H} and \ref{fig:H fit Ar4H} we show the results for the MD simulations of the time-dependent energy relaxations of both Ar and H$^+$ particles within Ar$_4$H$^+$ nascent cluster in comparisons with the theoretical predictions, indicated by the solid curves. The theoretical curves shown in Fig.\ref{fig:Ar fit Ar4H} or in Fig.\ref{fig:H fit Ar4H} have been computed with a single fitting parameter, the effective hard-sphere cross section, $\sigma$: $\sigma_{Ar}= 1.5*10^{-14}$cm$^2$ for Ar+Ar collisions, and $\sigma_{H^{+}}= 2.4*10^{-14}$ cm$^2$ for Ar+H$^+$ collisions. These parametric cross sections are in a general agreement with the quantum cross sections computed for collisions of free particles\cite{AVPhelps_2000,Phelps1992}. In the numerical calculations of the kinetic energy relaxation of Ar$_n$H$^+$ ($n \leq 4)$) clusters, $Q_n(\epsilon_k,t)$ can be ignored because the lifetime of these clusters are mostly determined by their cooling rates.

The excellent agreement between theoretical predictions and results of simulations indicates that growth of ultra-small clusters ($n\leq 4$) with the open or closed first atomic shell occurs mostly as the internal kinetic energy of cluster particles have been cooled enough to increase the Ar sticking probability in Ar$_n$H$^+$ +Ar collisions.

Scenario 2. \textit{The fast growth of larger clusters: effect of outer cluster shells}. The processes of internal energy relaxation and cluster growth can be substantially altered if the rate of cluster growth surpasses that of the cluster cooling. This phenomenon manifests in sufficiently large Ar$_n$H$^+$ clusters, wherein the interactions of outer or inner cluster shells with the surrounding bath gas atoms differ. Ar atoms in outer shells have significantly smaller binding energies and this creates favorable conditions for the redistribution of kinetic energy of incoming Ar atoms over various internal degrees of freedom and increasing the probability of $n\rightarrow n+1$  sticking collisions. The fast dynamics of the energy relaxation and growth of the clusters with outer shells ($n\geq 5$), i.e. shorter life-time of Ar$_n$H$^+$ clusters, are correlated with a presence of their long-range configurations (a) as it was discussed in Section \ref{sec::Lifetimes}. The different relaxation rates of the long-range (a) and tight (b)configurations effect the time-dependence of the average kinetic energy computed  using the Boltzmann equation, and thus should not be neglected.

To include the distinction between the cooling  of inner and other cluster shells, the distribution function $f_n(\epsilon_k,t)$ in Eq.\ref{eq:Boltzmann} is formally split into two independent distribution functions $f_n(\epsilon_k,t)=  w_a*f_{n,a}(\epsilon_k,t)+ w_b*f_{n,b}(\epsilon_k,t)$, one for the long-range (a) and  tight (b) configurations discussed in Section.\ref{sec::Lifetimes}. The statistical weight $w_a$ and $w_b$ describe an initial populations of the long-range and tight configurations ($w_a + w_b=1$). The decay rates of the cluster populations with the long-range (a) or tight (b) configurations, i. e. the different dynamics of the cluster growth for each configuration, are given by the independent sink functions $Q_{n,a}$ and  $Q_{n,b}$:
\begin{eqnarray}
    Q_{n,i}= - f_{n,i}(\epsilon_k,t)/\tau_{n,i}(\epsilon_k),  ~ ~~~  i= a, b ,
\end{eqnarray}
where $\tau_{n,a}$ and $\tau_{n,b}$ are the life-times for the cluster  with the long-range and tight configurations respectively. Details of computation procedure for these distribution functions are given in the Appendix.

\begin{figure}[ht!]
    \includegraphics[scale=.266]{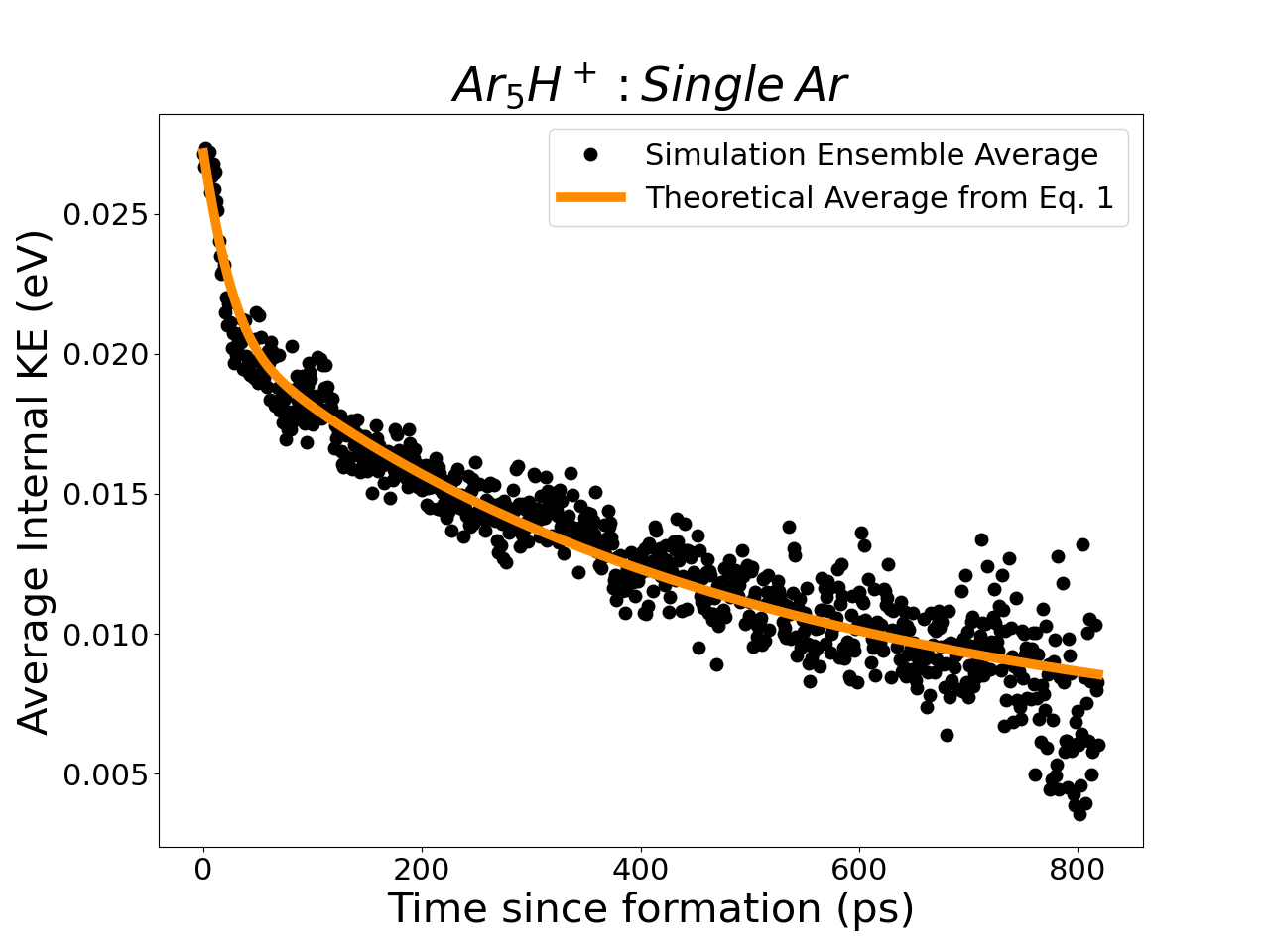}
    \caption{Shown is $\overline{\epsilon_{k,Ar}}$ inside an Ar$_5$H$^+$ cluster, from the ensemble of T=40K simulations, alongside the time-dependent average calculated from the Boltzmann equation with the presence of $Q_n(\epsilon_{k},t)$, and using the initial distribution function directly from our ensemble of simulations to initialize the (a) and (b) initial distributions.}
    \label{fig:Ar Fit Ar5H}
\end{figure}

\begin{figure}[ht]
    \includegraphics[scale=.266]{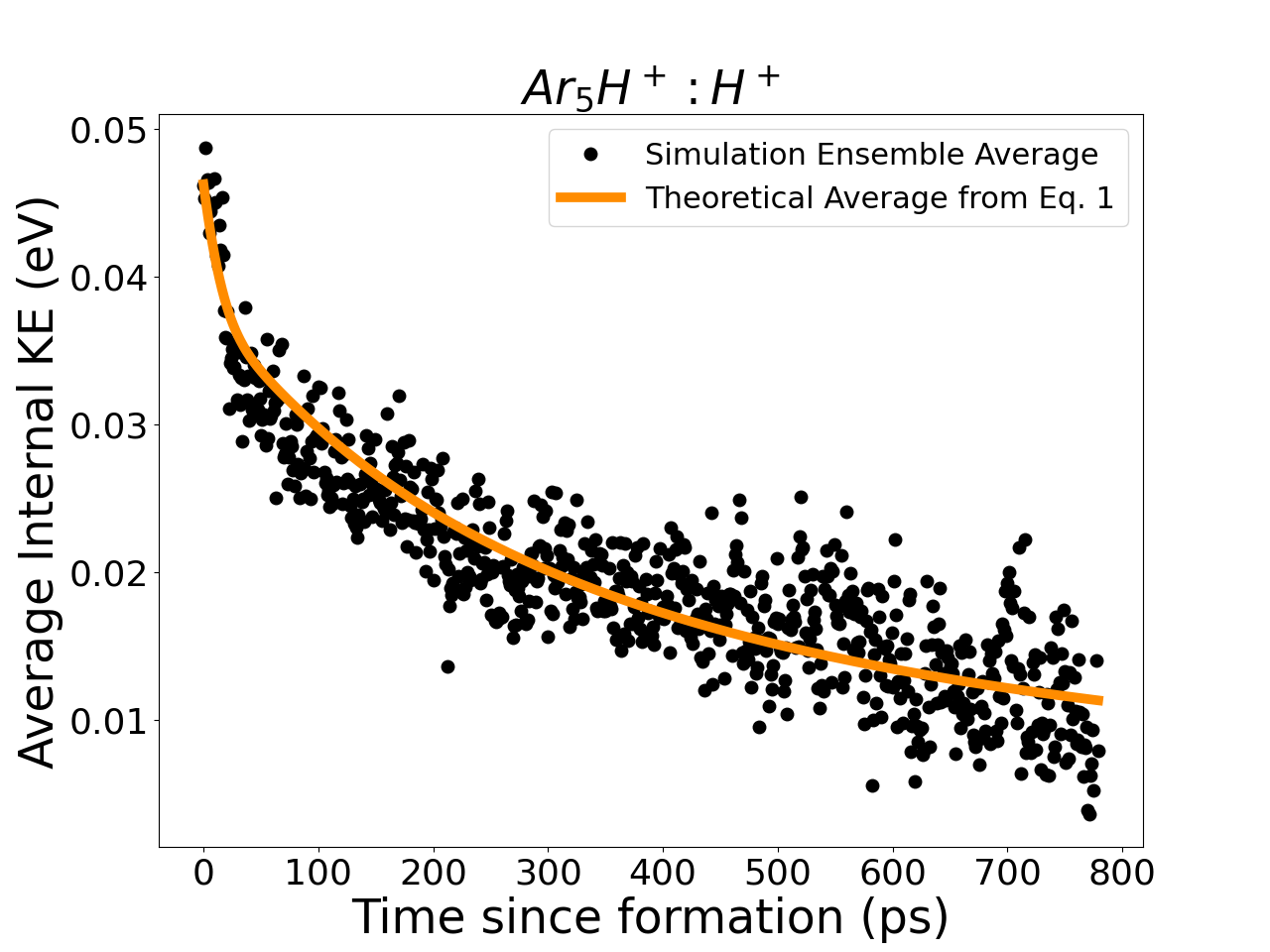}
    \caption{Shown is $\overline{\epsilon_{k,H^+}}$ inside an Ar$_5$H$^+$ cluster, from the ensemble of T=40K simulations, alongside the time-dependent average calculated from the Boltzmann equation with the presence of $Q_n(\epsilon_{k},t)$, using the initial distribution function directly from our ensemble of simulations to initialize the (a) and (b) initial distributions.}
    \label{fig:H Fit Ar5H}
\end{figure}

The cluster growth in $n\rightarrow n+1$ transitions requires a few collisions between Ar$_n^*$H$^+$ ($n\geq 5$) and free Ar atoms. Such collisions mostly involved clusters with the (a)-type of long-range configurations, which are predominately populated ($w_a$ significantly exceeds $w_b$). Complete decay of the long-range population occurs for the short time $\sim$ 50-70 ps. It involves a larger fraction of the Ar$_n^*$H$^+$ ensemble ($w_a > 60\%$), but does not influence strongly the averaged kinetic energies of the cluster Ar ($\overline{\epsilon_{k,Ar}}$) and H$^+$ ($\overline{\epsilon_{k,H^+}}$) particles. The energy-relaxation of the small fraction of the nascent clusters with the (b)-configuration similar to the relatively slow relaxation of Ar$_n^*$H$^+$ ($ n\leq 4$), except for highly excited metastable configurations. The empirical values for the life-time parameters $\tau_{n,a}$ and $\tau_{n,b}$ for different configurations allow to reach a very good agreement between theoretical predictions of the dynamics of the cluster cooling and results of the simulations. To illustrate dynamics of the  cooling clusters with inner and outer shells, we show in Figs.\ref{fig:Ar Fit Ar5H} and \ref{fig:H Fit Ar5H} the results of the MD simulations and theoretical analysis of the energy relaxation of Ar ($\overline{\epsilon_{k,Ar}}$) and H$^+$ ($\overline{\epsilon_{k,H^+}}$) particles in Ar$_5$H$^+$ clusters. 


The parameters required for the theoretical description of the Ar$_5$H$^+$ cluster cooling and growth processes via solutions of the Boltzmann equation are determined for Ar and H$^+$ cluster particles. For the Ar atom's cooling the following parameters were extracted: $\sigma_{Ar}=1.3 *10^{-14}$ cm$^2$, $\tau_{a,Ar}=23$ ps, and the energy dependent life-time of the $b$-type clusters reflects the decay of the clusters with the high kinetic energies $\tau_{b}(\epsilon_{k}) = \tau_b *(1+ exp[-(\epsilon_k-\epsilon_{c})/\epsilon_w]$. In the expression for $\tau_{b,Ar}$ the scaling time is $\tau_b = 15$ps and the  energy parameters $\epsilon_{c}=59$ meV corresponding to the kinetic energy threshold above that the tight the populations of the clusters quickly decays and $\epsilon_w $ describing the energy width of this threshold region. In our numerical calculations the width of the threshold areas was simply taken as $0.2$ $k_B$T, as it is described in the Appendix. The computational parameters for the energy relaxation of H$^+$ are:  $\sigma_{H^{+}}=2.8*10^{-14}$ cm$^2$,  $\tau_{a,H^+}=17$ ps, $\tau_{b,H^+}=11$ ps, and $\epsilon_{c,H^+}=100$ meV. The statistical weights of the long-range and tight configurations are $w_a \simeq 63\%$ and $w_b \simeq 27\%$ respectively. The proposed theoretical model recreates the sharp depopulation of the nascent Ar$_5^*$H$^+$ clusters in the first $\sim$50-80ps as it is shown in Fig.\ref{fig:NumPointsAr5H} and accurately describe the cluster cooling processes depicted in Figs.\ref{fig:Ar Fit Ar5H} and \ref{fig:H Fit Ar5H}. 

\section{\texorpdfstring{$Ar$ or $H^+$ Particles in a Single Permanently Growing Clusters} {Section:Sub-System}}

\begin{figure}[ht]
    \includegraphics[scale=.266]{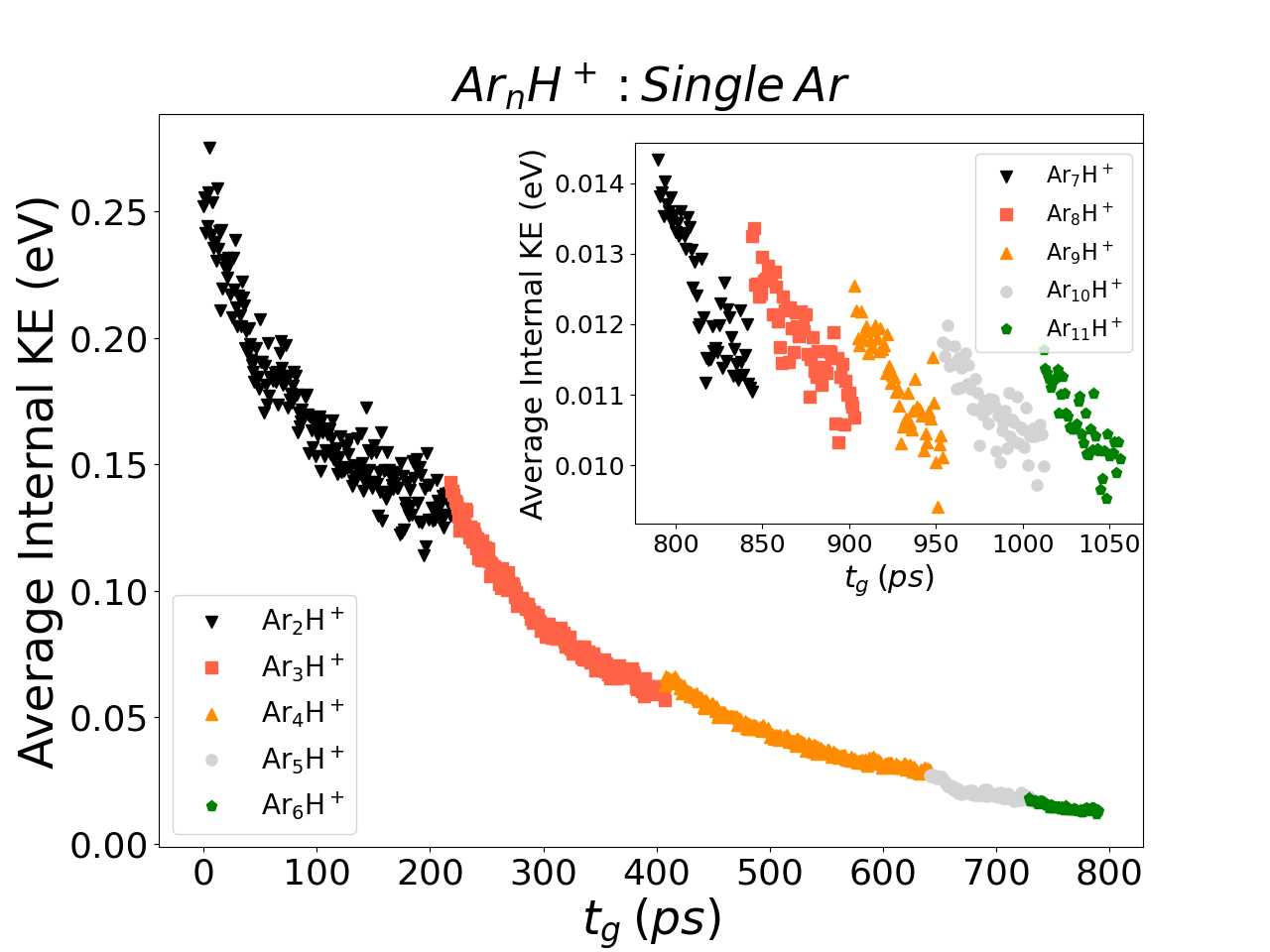}
    \caption{Shown is $\overline{\epsilon_{k,Ar}}$ inside an average Ar$_n$H$^+$ cluster, starting from $n=2$, from the ensemble of T=40K simulations. The different markers/colors represent different $n$, increasing from left to right across the figure ($n=2$ to $n=6$). While the inset shows the range of $n=7$ to $n=11$.}
    \label{fig:Ar Ave Stitched}
\end{figure}

\begin{figure}[ht]
    \includegraphics[scale=.266]{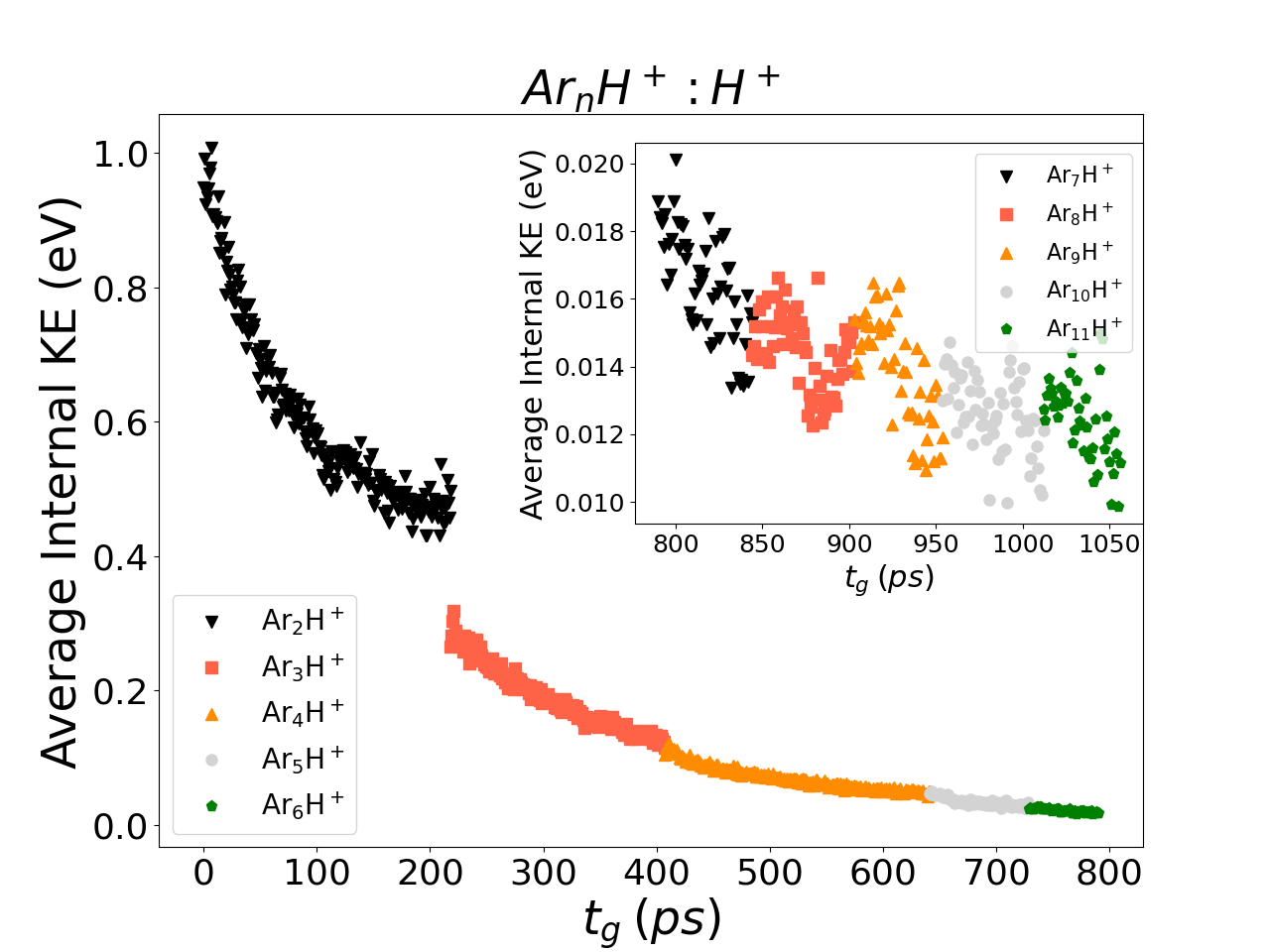}
    \caption{Shown is $\overline{\epsilon_{k,H^+}}$ inside an average Ar$_n$H$^+$ cluster, starting from $n=2$, from the ensemble of T=40K simulations. The different markers/colors represent different n, increasing from left to right across the figure($n=2$ to $n=6$). While the inset shows the range of $n=7$ to $n=11$.}
    \label{fig:H Ave Stitched}
\end{figure}

In previous sections, we described the relaxation dynamics in the ensemble of small clusters with specified sizes $n$. An alternative approach is to analyze parameters of Ar and H$^+$ cluster particles in a single permanently growing cluster as regular particles representing an embryo of a new phase. In this view, the Ar$_n$H$^+$ cluster, independently of their value of $n$, can be treated as an open subsystem that exchanges energy and particles with the equilibrium Ar gas\cite{OurArticle}. Thus, it is possible to randomly sample a single Ar atom or H$^+$ ion from the cluster phase at any time during the non-equilibrium cluster growth. Such sampling allows us to establish an averaged behavior of particles inside the new phase,i.e. the growing single Ar$_{n(t_g)}$H$^+$ cluster with increasing number of Ar atoms $n(t_g)$, where $t_g$ is the time since the beginning of the growth process. To model the averaged time evolution of the particle inside the growing embryo of new phase, we can use the averaged kinetic energies $\overline{\epsilon_{k,Ar}}$ and $\overline{\epsilon_{k,H^+}}$ inferred from the simulation of Ar$_{n}$H$^+$ ensembles with the specific $n$-values. We are suggesting, that the growing cluster gains particles after the average lifetime estimated in our simulations for the specific size $n$. According to this approach, the averaged kinetic energies $\overline{\epsilon_{k,Ar}}$ and $\overline{\epsilon_{k,H^+}}$ in a growing cluster are represented by data shown in Figs. \ref{fig:Ar Ave Stitched} and \ref{fig:H Ave Stitched} respectively.

There are few general trends in the data shown for the Ar and H$^+$ particles. When a cluster undergoes a non-thermal size transitions, $n \rightarrow n+1$, there may be a discontinuity in $\overline{\epsilon_{k,Ar}}$ and $\overline{\epsilon_{k,H^+}}$ values. This is expected for small clusters, as capturing free Ar atom(s) leads to the sharp excitation of the internal degrees of freedom in the cluster. These jumps in the kinetic energies typically decrease with increasing cluster size. Related discontinuity were also observed in the total internal kinetic energy of individual clusters, $\varepsilon_{k}$ as shown in Fig.\ref{fig:SingleClusterKE}. These discontinuities vanished when clusters reached the equilibrium with the ambient Ar gas (equilibration of the chemical potential of two phases). At this equilibrium stage, the energy of thermal fluctuations became comparable with values of energy jumps at $n\rightarrow n \pm 1$ transitions.

The data presented in Fig.\ref{fig:H Ave Stitched} indicates a notable reduction in the average H$^+$ kinetic energy during the cluster growth process from the $n=2$ to $n=3$ state. Specifically, the observed decrease in kinetic energy can be explained by efficient losing of the H$^+$ kinetic energy due to interaction with an additional Ar in the inner atomic shell. Processes of cooling and excitation of the cluster H$^+$ ions are significantly transformed with growth of the cluster size. In ultra-small clusters as well as small molecular ions, the ion can lose kinetic energy due to permanent efficient encounters with the heavy Ar atoms from the inner shell and also due to long range interaction with free Ar atoms from the bath gas. In the terms of molecular physics, these collisions can efficiently quench highly excited ro-vibrational modes involving motion of H$^+$ ion \cite{Stoecklin2016,Hansen2014}. For the large size clusters with inner and outer atomic shells, where the H$^+$ is trapped inside the Ar cage, the direct energy transfer collisions between H$^+$ and free Ar are less effective, and cooling of all cluster particles occurs mostly due to Ar+Ar collision involving the bath gas.

The data in Figs.\ref{fig:Ar Ave Stitched} and \ref{fig:H Ave Stitched} show the general cooling of the internal energy of the new phase (clusters) during non-equilibrium stage of the cluster growth. Randomly chosen Ar and H$^+$ cluster particles in the ensemble of independently growing clusters represent the relaxation dynamics of the new phase. Growth and cooling processes are compensated when the new phase reached the equilibrium at T=40K, and fluctuations of the cluster size and particle energies are described by the thermodynamic equilibrium between two phases, the free Ar gas and the Ar cluster atoms \cite{OurArticle}.

\section{Conclusion}

We conducted MD simulations to study of nucleation and non-equilibrium growth of nano-particles, focusing on three fundamental aspects of small cluster nucleation: (1) formation of nascent nano-clusters in highly excited rotational-vibrational states, (2) competition between cluster growth and internal energy relaxation, and (3) influence of the cluster shell structures and parameters of ambient gas on the cluster cooling and growth rates. Our conclusions were based on the data from simulations of Ar$_n$H$^+$ cluster nucleation initiated by H$^+$ ions in the cold metastable Ar gas and theoretical modeling of relaxation processes.

The time-dependent potential and kinetic  energies of all cluster particles within growing Ar$_n$H$^+$ clusters have been inferred from the results of MD simulations. These data were used to track the time-dependent energy relaxation and nucleation processes in the ensemble of independently growing Ar$_n$H$^+$ clusters. Our findings show that at low temperatures (T=40K) small nascent clusters with $n\leq 11$ are created in highly excited ro-vibrational states, i.e. with configurations which are far from equilibrium. Such type of excited states of the new phase is a common feature of the recombination and relaxation phenomena in plasma and atomic physics \cite{LLkinetics}.

Collisions of the cluster particles with the Ar bath gas atoms are responsible for both cooling and growing processes. Both processes can strongly influence each other and for the nano-size Ar$_n$H$^+$ clusters resulting rates of cooling and growth depend on the cluster shell structure, cluster size and the temperature and density of the ambient gas. As the Ar$_n$H$^+$ cluster grows, the scale of excitation of nascent clusters decreases with increasing $n$, and the growth process becomes more diffusive in nature.

It was also observed that within the small Ar$_n$H$^+$ ($n\leq 4)$ clusters, the kinetic energies of inner shell Ar atoms are quickly equalized on time scales much shorter than the typical collisional time between cluster and the bath gas atoms. The tight configurations of these very small clusters and large kinetic energies of inner particles require significant time to cool cluster's internal degrees of freedom and to provide necessary conditions for cluster growth such as an increase of the probability of sticking collisions of the bath gas atoms with cluster.

The Ar$_n$H$^+$ ($n\geq 5)$ cluster with the inner and outer atomic shells demonstrate different dynamics of the energy relaxation and growth. Small kinetic and potential energies of the outer shell Ar atoms in the long-range cluster configurations and their efficient interaction with the cold bath gas yield significant increase of the cluster growth rate, which becomes larger than rates of corresponding cooling collisions.

The results of the MD simulations of the cluster cooling and growth can be described by a simplified model of the cluster energy relaxation due to short range energy-transfer collisions between hot cluster particles and cool Ar atoms of the ambient gas. It was shown in the framework of such simple model, the numerical solutions of the Boltzmann equation for the energy relaxation of hot Ar and H$^+$ cluster particles in the cold Ar gas are in a very good agreement with the results of simulations for clusters with the inner Ar shell ($n\leq 4$). Theoretical description of the time-dependent energy relaxation and growth of Ar$_n$H$^+$ ($n\geq 5$) clusters also provides a good agreement with result of MD simulations, but it requires to employ empirical formulas for the rate of growth of the clusters with outer atomic shells. 

The data on cooling of the ensemble of Ar$_n$H$^+$ clusters with specific number $n$ of Ar atoms have been used to illustrate the averaged energy relaxation processes in the permanently growing cluster with the time-increasing number of atoms  $n=n(t_g)$.

The results of present work provide a deep inside information on the characteristics and intensity of relaxations processes responsible for the nucleation and growth on nano-size clusters in the cold atomic gas in a presence of ionized particles. We expect that some of obtained results on the non-equilibrium growth and energy relaxation in nano-size particles, such as increase of the internal energy of charged nano-particles during nucleation process under or an influence of the atomic/molecular shell structure on the growth rate, will be valid for different environmental conditions.

\section*{Acknowledgment}
V.K. acknowledges support from the National Science Foundation through a grant for the ITAMP at the Harvard-Smithsonian Center for Astrophysics; R.C. was supported by
the National Science Foundation (NSF), Grant No. PHY-2034284.

\appendix*
\section{}
Cooling of the internal energy of nascent clusters predominately occurs in short-range binary collisions of the free Ar atoms with cluster atomic particles. These short-time and short-range encounters are mostly responsible for changes in the kinetic energies of colliding particles and can be considered as elemental steps of the internal energy relaxation of nascent Ar$_n$H$^+$ clusters. The rate of the kinetic energy-transfer process $\epsilon'_k\rightarrow \epsilon_k$ in short-range collisions between the hot cluster particles and Ar atoms of the  bath gas can be described by the kernel of the Boltzmann kinetic equation $B(\epsilon_k|\epsilon'_k)$ expressed via the anisotropic differential cross sections \cite{KHARCHENKOFastnNitrogen,SXeRelaxation}. Relatively expansive numerical calculations are required to establish $B(\epsilon_k|\epsilon'_k)$'s functional form, but significant simplification in the solution of the Boltzmann equation can be achieved, if the energy and angular dependence of the cross sections are neglected \cite{AndersonHardSphere,KHARCHENKOFastnNitrogen}. This replacement of real cross sections with the hard sphere provides rather accurate description of the time-evolution of energy distribution functions of hot particles in atomic and molecular collisions. The values of single empirical parameter of the hard sphere approximation, the cross section $\sigma$, are usually close to values of corresponding momentum-transfer cross sections.

To efficiently solve Eq.\ref{eq:Boltzmann}, we transform it into a unitless equation, where unitless energy $e$ is measured in multiples of $k_BT$, with $k_B$ as the Boltzmann constant, $e=\epsilon_k/k_BT$. The unitless time is given by $\tau = n_b \sigma v_T t $ where $n_b$ is the density of the bath gas, $\sigma$ is the HSA cross section, $v_T$ = $\sqrt{2k_BT/m}$ is the scaling thermal velocity, and $m$ is the mass of the hot cluster particle which is the Ar or H$^+$. Unitless kinetic energies of the cluster particles before and after a single collision are:
\begin{equation}
  \label{eq:1}
  e_{\mathrm{in}} \equiv \frac{\epsilon_{k,\mathrm{in}}}{k_BT} ,
  \quad e_{\mathrm{out}} \equiv \frac{\epsilon_{k,\mathrm{out}}}{k_BT} .
\end{equation}
The ratio of the hot particle mass $m$ to the mass $m_b$ of Ar bath gas atoms, $\kappa = m/m_b$, is a fundamental parameter of the energy relaxation kinetics \cite{AndersonHardSphere} regulating the rate of momentum or energy transfer in the binary collisions.

The unitless kernel of the Boltzmann equation $b(e_{\mathrm{out}}|e_{\mathrm{in}})$ can be written as \cite{AndersonHardSphere,KHARCHENKOFastnNitrogen}: 

 \begin{widetext}
  \begin{eqnarray}
  b(e_{\mathrm{out}}|e_{\mathrm{in}}) =
    \frac{(\kappa + 1)^2}{8 \kappa \sqrt{e_{\mathrm{in}}}} \times 
  \left\{
    \begin{array}{ll}
      e^{(e_{\mathrm{in}} - e_{\mathrm{out}})}
      \left[
      erf\left(\frac{(\kappa+1)\sqrt{e_{\mathrm{in}}} +
      (\kappa - 1) \sqrt{e_{\mathrm{out}}}}{2 \sqrt{\kappa}}\right) -
      erf\left(\frac{(\kappa+1)\sqrt{e_{\mathrm{in}}} -
      (\kappa - 1) \sqrt{e_{\mathrm{out}}}}{2 \sqrt{\kappa}}\right)
      \right] +
      &  \\
      erf\left(\frac{(\kappa-1)\sqrt{e_{\mathrm{in}}} +
      (\kappa + 1) \sqrt{e_{\mathrm{out}}}}{2 \sqrt{\kappa}}\right) -
      erf\left(\frac{(\kappa-1)\sqrt{e_{\mathrm{in}}} -
      (\kappa + 1) \sqrt{e_{\mathrm{out}}}}{2 \sqrt{\kappa}}\right)
      & \\
      \text{if } e_{\mathrm{in}} > e_{\mathrm{out}}, &  \\
      & \\
      erf\left(\frac{(\kappa-1)\sqrt{e_{\mathrm{in}}} +
      (\kappa+1)\sqrt{e_{\mathrm{out}}}}{2 \sqrt{\kappa}}\right) -
      erf\left(\frac{-(\kappa-1)\sqrt{e_{\mathrm{in}}} +
      (\kappa+1)\sqrt{e_{\mathrm{out}}}}{2 \sqrt{\kappa}}\right) +
      &  \\
      e^{(e_{\mathrm{in}} - e_{\mathrm{out}})}
      \left[
      erf\left(\frac{(\kappa+1)\sqrt{e_{\mathrm{in}}} +
      (\kappa-1)\sqrt{e_{\mathrm{out}}}}{2 \sqrt{\kappa}}\right) -
      erf\left(\frac{-(\kappa+1)\sqrt{e_{\mathrm{in}}} +
      (\kappa-1)\sqrt{e_{\mathrm{out}}}}{2 \sqrt{\kappa}}\right)
      \right]
      &  \\
      \text{if } e_{\mathrm{in}} < e_{\mathrm{out}}. & 
    \end{array}
  \right.
    \end{eqnarray}
\end{widetext}
where $erf(x)$ is the error function. This transforms Eq.\ref{eq:Boltzmann} into the following:
\begin{equation}\label{eq::DimensionlessBoltzmann}
    \frac{\partial}{\partial \tau}f(e,\tau) = \int b(e|e')f(e',\tau)de' - f(e,\tau)\int b(e'|e)de' .
\end{equation}
 The sink/source term $Q(e,\tau)$  doesn't influence the cooling rates of Ar$_n^*$H$^+$ clusters with $n\leq 4$ and it is omitted  in Eq.\ref{eq::DimensionlessBoltzmann} for simplicity.
 
For the numerical solution of Eq.\ref{eq::DimensionlessBoltzmann}, the energy of the hot particle is discretized with the $N$ small steps $\delta e$, where $\delta e \ll 1$. In our solution, the unitless energies $e$ and $e'$ have been discretized with the bin width $\delta e = 0.01 $, i.e. the discrete step in calculations of the real kinetic energies was 0.01$k_BT$. The small value of $\delta e$ has been selected to verify an accurate convergence of numerical solutions of Eq.\ref{eq::DimensionlessBoltzmann} to the Maxwell distribution as $\tau \rightarrow \infty$. The number of bins $N$ can be different for the clusters of different sizes $n$. The discrete kinetic energy distribution functions of the Ar and H$^+$ particles of the nascent clusters were directly inferred from results of simulations. These distributions have been used as the initial distribution functions in solutions of Eq.\ref{eq::DimensionlessBoltzmann}. 

In the discrete representation, the kinetic energy distribution functions of hot cluster particles $f(e,\tau)$ are given by the $N$-dimensional vector $ \vec{\mathsf{f}}(\tau)$. The kernel of the discretized kinetic equation is represented by the tensor $\mathbf{W}_{ij}$, the N x N matrix. Shown are the formal discretizations of energy, $ \vec{\mathsf{f}}(\tau)$, and Eq. \ref{eq::DimensionlessBoltzmann}:
\begin{equation*}
  \vec{\mathsf{e}} = \left[ \frac{\delta e}{2}, \; \frac{3 \delta e}{2}, \;
  \frac{5 \delta e}{2}, \ldots, \; e_{\mathrm{max}} - \frac{\delta
    e}{2} \right]  , \quad \delta e = \frac{e_{\mathrm{max}}}{N} ,
\end{equation*}
\begin{equation*}
  \vec{\mathsf{f}}(\tau) = \left[ f(e_1, \tau), \; f(e_2, \tau), \;
    \ldots, \; f(e_N, \tau) \right] ,  
\end{equation*}
\begin{equation*}
  w_{ij} = b(e_i | e_j) \, \delta e
\end{equation*}
\begin{eqnarray}   
  \frac{\mathrm{d}}{\mathrm{d} \tau} f_i &&=
  \sum_{j=1}^{N} w_{ij} \, f_j  -  f_i \sum_{j=1}^{N} w_{ji} \nonumber\\
  &&=\sum_{j=1}^{N} \left(w_{ij} - \delta_{ij} \sum_{j=1}^{N} w_{ji}\right) \, f_j \nonumber\\ 
  &&=\sum_{j=1}^{N} W_{ij} \, f_j  \nonumber
\end{eqnarray}
\begin{equation}\label{eq::SolvBoltzmann}
  \frac{\mathrm{d}}{\mathrm{d}\tau} \, \vec{\mathbf{f}} =
  \mathbf{W} \, \vec{\mathbf{f}}, \qquad  
  \mathbf{W}_{ij} \equiv w_{ij} - \delta_{ij} \sum_{j=1}^{N} w_{ji} .
\end{equation}
The formal solution of Eq.\ref{eq::SolvBoltzmann} for the vector $\vec{\mathbf{f}}(\tau)$ was obtained using   the evolution operator $\rm{exp}[ \tau \,\mathbf{W} ] $ \cite{matrixexp2003,higham2005}:
\begin{equation}
  \vec{\mathsf{f}}(\tau) = \rm{exp}[ \tau \,\mathbf{W}  ] \, \vec{\mathbf{f}}(0) ,
\end{equation}
where $\mathbf{f}(0)$ \cite{HydrogenRelaxation} the initial energy distribution function. The computed vector functions $ \vec{\mathsf{f}}(\tau)$ has been converted back to the distribution function $f_{n}(\epsilon_k,t)$ of the kinetic energies of hot particles within Ar$_n$H$^+$ clusters, as it is written in Eq.\ref{eq:Boltzmann}.

The obtained time-dependent distribution functions $f_{n}(\epsilon_k,t)$ have been employed for calculations of the time-dependent averaged energy $\langle \epsilon_{k,n}(t) \rangle$ of the Ar$_n^*$H$^+$ 
 ($n\leq 4$) cluster particles:
\begin{equation}
    \overline{\epsilon_{k,n}}  = \int f_n(\epsilon_k,t)  \epsilon_k d \epsilon_k,
\end{equation}

The results of these calculations were used for analysis of cooling and growing processes in nano-clusters with $n \leq 4$ as it shown in  Figs.\ref{fig:Ar fit Ar4H} and \ref{fig:H fit Ar4H}  for Ar and H$^+$ cluster particles.

The significant difference in the cluster growth (Figs.\ref{fig:NumPointsAr4H} and \ref{fig:NumPointsAr5H}) and cooling processes (Figs.\ref{fig:Ar fit Ar4H} and \ref{fig:Ar Fit Ar5H}) can be seen for clusters with the inner $ n\leq 4$ and outer $ n\geq 5$ shells.
Computation of the time-dependent energy distribution functions of hot particles inside rapidly growing clusters with the outer atomic shell(s) requires a consideration of the decay of the nascent  Ar$_n^*$H$^+$ ($n\geq5$) population due to relatively fast $n \rightarrow n\pm1$ transitions.
 Thus, the presence of the sink function, $Q_n(\epsilon_k,t)$, cannot be ignored in the numerical analysis of the cooling processes for clusters with $n\geq 5$. 
 
The configurations of the nascent Ar$_n$H$^+$ excited clusters can be formally split into two categories described in the Section VI: (a) the long-range configurations with low kinetic and potential energies of the outer shell Ar atoms interacting effectively  with the Ar bath gas; and (b) the tight configurations  where outer shell Ar atoms with significant kinetic energies are trapped close to the inner shell. These configurations are characterised by the different relaxation processes and time-dependent distribution functions $f_{n,a}(\epsilon_k,t)$ and $f_{n,b}(\epsilon_k,t)$. The statistical weights $w_a$ and $w_b$  ($w_a +w_b =1$) of these configurations in the ensemble of nascent Ar$_n$H$^+$ cluster can be inferred form the results of simulations, but in the present modeling $w_a$ has been employed as the fitting parameter. For example, we evaluated  from the fitting of the simulation results  that $w_a \simeq 0.63 $ and $w_b \simeq 0.37$ for the ensemble of nascent Ar$_5^*$H$^+$ clusters. The sink function for the $f_{n,i}(\epsilon_k,t)$ distributions ($i=a,b$) can be written as $Q_{n,i} = -f_{n,i}(\epsilon_k,t)/\tau_{n,i}(\epsilon_k)$, where $\tau_{n,i}(\epsilon_k)$ are the  scaling life-time for the clusters with the long-range or tight configurations. 

The population of the (a)-type of Ar$_n^*$H$^+$ clusters with the long-range configurations and relatively small kinetic energies decays very quickly due to $n \rightarrow n+1$ growth transitions involving collisions between the outer shell Ar atoms and cold atoms of the bath gas. These transitions occur very quickly and the dependence of the life-time $\tau_{n,a}(\epsilon_k)$ from the kinetic energies of outer shell Ar atoms can be neglected: $\tau_{n,a}(\epsilon_k) \simeq \tau_{n,a}$.  
    
The life-time of the clusters with the tight  configurations (b) can depend on the kinetic energies of the cluster particles. Although the potential energies, i.e. configurations of the (b)-type clusters, are close to the equilibrium values of Ar$_n$H$^+$, these excited clusters have been created with the relatively high kinetic energies in the direct collisions between free Ar atoms and cluster core, the cluster particles from the inner shell. The relaxation processes for the (b)-type tight configurations  are similar to the relaxation of the tight clusters with inner shell only ($n\leq 4$). The life-time $\tau_{n,b}(\epsilon_k)$ can be relatively long because the high kinetic energies of cluster particles and relatively slow cooling processes. On the other hand, the presence of highly energetic cluster particles with the kinetic energies above some threshold $\epsilon_{c,n}$ can lead to redistribution of the internal kinetic energy inside cluster and following detachments of the weakly bound Ar atoms from the outer shells, i.e. decay of the Ar$_n^*$H$^+$ cluster population. We have used  empirical expression for the life-time of the clusters with the tight (b) configurations, which describe the discussed above life-time behavior:
\begin{equation}
    \tau_{b}(\epsilon_{k}) = \tau_b *(1+ exp[-(\epsilon_k-\epsilon_{c,n})/\epsilon_w]), 
\end{equation}
where $\epsilon_{c,n}$ is the empirical kinetic energy threshold for the  stability of clusters with the outer shells, $\tau_{b}$ is the empirical scaling time and $\epsilon_w$ describes the energy width of the threshold region, which was taken as 0.2 $k_B$T. This value was chosen such that $1/\tau_b(\epsilon_{k})$ will act similar to step function with narrow width around the threshold energy. The last equation shows that the energy dependent life-time of the high energy  fraction of $f_{n,b}(\epsilon_k,t)$ is relatively short, but for the bulk of the $f_{n,b}(\epsilon_k,t)$ distribution the decay process is slower than the typical time required for the energy relaxation. The relaxation of these long-living configurations was shown similar to the relaxation of Ar$_n$H$^+$ ($n\leq 4$) clusters, and therefore their decay can also be neglected in solving the Boltzmann kinetic equation.

\bibliography{references}
\end{document}